\documentclass[final,1p,times]{elsarticle}
\journal{Nuclear Physics B}
\usepackage{graphicx}
\usepackage{amsmath}
\usepackage{amssymb}
\usepackage{dsfont}
\usepackage{wasysym}
\usepackage{dcolumn}
\usepackage{bm}

\begin{document}
\begin{frontmatter}



\title{Exact solutions of a one-dimensional mixture of spinor bosons and
spinor fermions}
\author[hk]{Shi-Jian Gu}
\author[hk,b]{Junpeng Cao}
\author[b]{Shu Chen}
\author[hk]{Hai-Qing Lin}
\address[hk]{Department of Physics and ITP, The Chinese
University of Hong Kong, Hong Kong, People's Republic of China}
\address[b]{Beijing National Laboratory for Condensed Matter
Physics, Institute of Physics, Chinese Academy of Sciences,
Beijing 100190, People's Republic of China}

\begin{abstract}
The exact solutions of a one-dimensional mixture of spinor bosons and spinor
fermions with $\delta$-function interactions are studied. Some new sets of
Bethe ansatz equations are obtained by using the graded nest quantum inverse
scattering method. Many interesting features appear in the system. For
example, the wave function has the $SU(2|2)$ supersymmetry. It is also found
that the ground state of the system is partial polarized, where the fermions
form a spin singlet state and the bosons are totally polarized. From the
solution of Bethe ansatz equations, it is shown that all the momentum, spin
and isospin rapidities at the ground state are real if the interactions
between the particles are repulsive; while the fermions form two-particle
bounded states and the bosons form one large bound state, which means the
bosons condensed at the zero momentum point, if the interactions are
attractive. The charge, spin and isospin excitations are discussed in
detail. The thermodynamic Bethe ansatz equations are also derived and their
solutions at some special cases are obtained analytically.
\end{abstract}

\end{frontmatter}


\section{Introduction}

Recently, tremendous experimental progresses have been taken in
the research of the one-dimensional (1D) trapped cold atoms \cite
{MGreiner01,AGolitz01,FSchreck01,HMoritz03,
h,b1,TKinoshita04,BParedes04,HMoritz05,SAubin05,101,102}. By means
of either magnetic or optical traps, the cold atom gas has been
realized. With Feshbach resonance, the scattering length and thus
the couplings among atoms can be manipulated. In addition, with
laser beams, one can confine particles in the valleys of periodic
potential of the optical lattice. These experimental tools
provide a platform to study some controllable condensed matter
systems. Theoretical, many methods have been applies to study
these problems. For example, the Gross-Pitaviskii (GP) theory is
widely adopted in dealing with the systems of Bose-Einstein
condensations with weak interactions. However, the GP theory is
based on a mean field approximation and has many shortcomings.
The boson-Hubbard model \cite{boson1,boson2} are used to describe
the low-energy physics of ultracold dilute gas of bosonic atoms
in an optical lattice. This model can not be solved exactly and
one has to use the numerical approaches or approximate methods
such as the valence bond mean field \cite{zhang}. Exactly
solvable models play an important role in the investigation of 1D
interacting many-particle systems. The exact solutions can supply
some believable results thus serve as a very good starting point
to understand the new phenomena and new quantum states in trapped
cold atomic systems.

Using laser beams or Feshbach resonance, the local interactions can be
strongly enhanced. The low-energy behavior of trapped cold atoms can be well
described by a system with $\delta$-function potentials. Fortunately, this
kind of problems can be solved exactly. The exact solutions of scalar bosons
with $\delta$-function interactions are obtained by Lieb and Liniger \cite%
{EHLieb63}. Li, Gu, Ying and Eckern study the two-component bosons, where
the intrinsic degrees of freedom (isospin) of bosons satisfy the SU(2)
symmetry \cite{YQLi03}. They find that the ground state of the system is not
an isospin singlet, but a polarized or ferromagnetic state. In fact, the
Bethe ansatz equations of multi-component bosons have already been obtained
by Zhou \cite{zhou}. Naturally, one should consider the effects of spin
exchanging interactions if the bosonic atoms have non-zero spin. An exactly
solved model of bosons with spin-1 is proposed in the Ref. \cite{cao}, where
the exchange interactions are embodied in the model Hamiltonian and the
system is spin-dependent. More physical properties of bosonic cold atomic
systems with $\delta$-function potentials can be found in the Refs. \cite%
{fuchs,liebseir,103,104,105,106}.

In experiments, atoms with inner degrees of freedom (hyperfine spin) are
prepared by catching several hyperfine sublevels of atoms. The spins can be
polarized by the external magnetic fields, thus it is easier to capture the
inter state of the fermions by using the laser beams. Now, people can
control the fermionic atoms steadily staying on some special inter states,
and realize the fermionic atoms with multi-component hyperfine spin \cite%
{HMoritz05}. The theoretical model of the spin-1/2 fermions with $\delta$%
-function interactions is solved by Gaudin \cite{gau} and Yang \cite%
{CNYang67}. Sutherland generalizes the results to the multi-component
fermionic systems \cite{suth}.

Most recently, the study of ultracold Bose-Fermi mixtures become a
remarkable research topic for many new matter phases may arise in these
systems. Experimenters have succeeded in preparing the mixtures of $^7$Li-$%
^6 $Li, $^{23}$Na-$^6$Li or $^{87}$Rb-$^{40}$K in the optical lattices \cite%
{16,17,18,19,20,21,4,5,6}. For example, a stable bosonic $^{87}$Rb and
fermionic $^{40}$K mixture in three dimensional optical lattices has been
realized \cite{5,6}. Theoretically, Lai and Yang obtain the exact solutions
of 1D mixture of spin-1/2 fermions and scalar bosons with the $\delta$%
-function interactions \cite{CKLai71}. They calculate the ground state
energy and gapless fermionic excitations in the thermodynamic limit \cite%
{CKLai72}. For more studies on the boson-fermion mixtures in the optical
lattices, please see the Refs. \cite%
{HFrahm05,MTBatchelor06,AImambekov06,ZXHu06}.

It is natural that one should consider the mixtures of
multi-component fermions and multi-component bosons. In this
case, both the fermions and the bosons have the intrinsic degrees
of freedom. It is well-known that the ground state of bosonic
systems with intrinsic degrees of freedom can be surprisingly
different from that of the scalar bosons. Furthermore, the ground
state of the spin-1/2 fermionic system with $\delta$-function
interactions is spin singlet, while the ground state of bosonic
system is isospin polarized or ferromagnetic state. One may
wonder what will happen if we mix the bosons and fermions both
with some intrinsic degrees of freedom? What is the new quantum
state in the mixtures? These issues are quite interesting and
important nowadays due to the rapid progress in the field of cold
atomic physics.

In this paper, we study a mixture of two-component fermions and
two-component bosons with $\delta$-function interactions. Because
the wave function of the system is supersymmetric and satisfies
the $SU(2|2)$ Lie superalgebra, we use the super or graded nest
quantum inverse scattering method
\cite{r1,zhao,eks,eks0,eks1,eks2,r2} to derive the exact solutions
of the system at 1D. We obtain the Bethe ansatz equations with
different gradings. We find that the ground state of the system
is partial polarized. If the interactions are repulsive ($c>0$),
all the momentum, spin and isospin rapidities in the solutions of
Bethe ansatz equations are real. If the interactions are
attractive ($c<0$), the Bethe ansatz equations may have the
complex solutions, where the fermions form the two-particle
bounded states and the bosons are condensed at the zero momentum
point at the ground state. We then study the low-lying
excitations such as charge, spin and isospin excitations in
detail. We also obtain the thermodynamic Bethe ansatz equations
at finite temperatures and find their analytic solutions at some
special limiting cases.

The integrable $SU(2|2)$-supersymmetry is a very interesting
issue. The Bethe ansatz for the corresponding quantum spin chain
is obtained by Kulish \cite{r1}, and the continue limit and low
energy behaviors are studied by Saleur \cite{r2}. In this paper,
we study the corresponding continue quantum gas model, which also
has many applications in the systems of cold atoms with hyperfine
structure. The low-lying excitation spectrum can be measured by
the scattering of phonons in experiments. Meanwhile, the cold
atoms with different internal states can be prepared in
experiments. The phonons scattering experiments on the bosonic
cold atoms have been done several years ago \cite{pho1,pho2}.
Thus the motivation of this paper is to give a prediction on the
ground state and low-lying excitation properties of the
bose-fermi mixture in the cold atom systems, for the excitations
in spin and charge sectors can show different behaviors.

The paper is organized as follows. In section 2, we introduce the
supersymmetry of the system and the generators of the
corresponding Lie superalgebra. In section 3, we derive the exact
solutions of the system by using the generalized quantum inverse
scattering method. We give the Bethe ansatz equations with the
BBFF grading, which are the foundations of our discussions. The
ground state properties are discussed in section 4 and the
low-lying excitations are discussed in section 5. The
thermodynamic Bethe ansatz equations are calculated in section 6
and some useful limit cases are discussed in section 7. Section 8
contains some conclusions and discussions.

\section{Supersymmetry of the system}

We study a 1D cold atomic system mixed by $N_{b_1}$ bosons of species $1$, $%
N_{b_2}$ bosons of species $2$, $N_{f_1}$ fermions of species $1$ and $%
N_{f_2}$ fermions of species $2$. The Hamiltonian reads
\begin{eqnarray}
H= \int_0^L \sum_{\alpha}\partial _x \phi_{a}^{\dag} (x) \partial_{x}
\phi_{a} (x) d x + \int_0^L \sum_{\alpha\beta} g_{ \mathrm{\alpha\beta}}
\phi_{\alpha}^{\dag} (x)\phi_{\beta}^{\dag}(x)
\phi_{\beta}(x)\phi_{\alpha}(x) d x,  \label{h1}
\end{eqnarray}
where $\alpha, \beta=b_1, b_2, f_1, f_2$ and $\phi_{\alpha}$ are the bosonic
or fermionic field operators. The bosonic field operators satisfy the
commutation relations, $[\phi_{\alpha}^{\dag}(x), \phi_{\beta}(y)]=
\delta_{\alpha\beta}\delta_{xy}$, while the fermionic field operators
satisfy the anticommutation relations, $\{\phi_{\alpha}^{\dag}(x),
\phi_{\beta}(y)\} =\delta_{\alpha\beta}\delta_{xy}$. According to the Pauli
exclusion principle, the $\alpha$ and $\beta$ in Hamiltonian (\ref{h1}) can
not be the same species of fermions. In this paper, we use the periodic
boundary conditions. The wave function of the system (\ref{h1}) is
supersymmetric, $\Psi(x_j, x_l)=P_{jl}\Psi(x_l, x_j)$, where $P_{jl}$ means
exchanging both the coordinates and the spins (isospins) of two particles $j$
and $l$. $P_{jl}=1$ for bosons or bosons and fermions thus the wave function
is symmetric, while $P_{jl}=-1$ for fermions thus the wave function is
antisymmetric.

The supersymmetry of the system (\ref{h1}) can be described by
the $SU(2|2)$ Lie superalgebra. The superalgebra $SU(2|2)$ has
fifteen generators and eight of them are fermionic
\cite{eks,eks0,eks1,eks2}. Moreover, the pure two-component
fermionic subsystem has the $SU(2)$ invariance and the pure
two-component bosons also has the $SU(2)$ invariance. In order to
explain the symmetry of the system more clearly, we introduce the
particle creation (annihilation) operators as
$a_{\sigma}^\dagger(x)$ ($a_{\sigma}(x)$), where
we assume two species of fermions carrying the different spins $%
\sigma=\uparrow, \downarrow$. There are four kinds of states at a given
position $x$,
\begin{eqnarray}
|0\rangle_x, \quad |\uparrow \rangle_x=a^{\dagger
}_{\uparrow}(x)|0\rangle_x, \quad |\downarrow \rangle_x=a^{\dagger
}_{\downarrow}(x)|0\rangle_x, \quad |\uparrow \downarrow \rangle_x=
a^{\dagger }_{\downarrow}(x)a^{\dagger }_{\uparrow}(x)|0\rangle_x.
\end{eqnarray}
The state $|0\rangle_x$ is vacuum and the state $|\uparrow \downarrow
\rangle $ represents that an atom-pair is localized on a single energy
level. Now, we introduce the generators of the superalgebra $SU(2|2)$. The
spin operators are defined as
\begin{eqnarray}
&&S^+= \int_0^L a_{\uparrow}^\dagger(x)a_{\downarrow}(x)dx, \quad
S^-=
\int_0^L a_{\downarrow}^\dagger(x)a_{\uparrow}(x)dx,  \nonumber \\
&&S^z=\frac{1}{2} \int_0^L
[a_{\uparrow}^\dagger(x)a_{\uparrow}(x)-a_{\downarrow}^\dagger(x)a_{%
\downarrow}(x)]dx.
\end{eqnarray}
The spin operators $S^+$, $S^-$ and $S^z$ form the $SU(2)$ Lie
algebra, where the commutation relations between the generators
are $[S^-, S^+]=2S^z, [S^+, S^z]=S^+, [S^-, S^z]=-S^-$. The above
spin operators are grassmann even (bosonic). We introduce the
pairing operators, which are also bosonic generators,
\begin{eqnarray}
&&\eta^+= \int_0^L
a_{\downarrow}^\dagger(x)a_{\uparrow}^\dagger(x)dx,
\quad \eta^-= \int_0^L a_{\downarrow}(x)a_{\uparrow}(x)dx,  \nonumber \\
&&\eta^z=\frac{1}{2} \int_0^L
[1-a_{\uparrow}^\dagger(x)a_{\uparrow}(x)-a_{\downarrow}^\dagger(x)a_{%
\downarrow}(x)]dx.
\end{eqnarray}
The pairing operators $\eta^+, \eta^-, \eta^z$ also form a $SU(2)$
Lie algebra with the commutation relations $[\eta^-,
\eta^+]=2\eta^z, [\eta^+, \eta^z]=\eta^+, [\eta^-,
\eta^z]=-\eta^-$. The eight fermionic generators are
\begin{eqnarray}
&&Q_{\sigma}=\int_0^L [1-a^\dagger_{\bar \sigma}(x)a_{\bar \sigma}(x)]
a_{\sigma}(x), \quad Q_{\sigma}^\dagger=\int_0^L [1-a^\dagger_{\bar
\sigma}(x)a_{\bar \sigma}(x)] a^\dagger_{\sigma}(x),  \nonumber \\
&&{\tilde Q}_{\sigma}=\int_0^L a^\dagger_{\bar \sigma}(x)a_{\bar \sigma}(x)
a_{\sigma}(x), \quad\quad\quad {\tilde Q}_{\sigma}^\dagger=\int_0^L
a^\dagger_{\bar \sigma}(x)a_{\bar \sigma}(x) a^\dagger_{\sigma}(x),
\end{eqnarray}
where $\bar \sigma$ means the spin with opposite direction of $\sigma$. The
fermionic operators are grassmann odd. These operators together with the
unit operator $\int_0^L 1 dx = L $ generate the Lie superalgebra $SU(2|2)$.

\section{Bethe ansatz solutions of the system}

In the following, we consider the case that all the coupling parameters are
equal, $g_{\alpha\beta}=c$. The system (\ref{h1}) has several integrable
lines. If $N_{b_2}=N_{f_1}=N_{f_2}=0$, the system degenerates to the scalar
bosons with $\delta$-function interactions which is solved by Lieb and
Liniger \cite{EHLieb63}. If $N_{f_1}=N_{f_2}=0$, the system degenerates to
the two-component $SU(2)$ bosons and is studied by Li, Gu, Ying and Eckern
\cite{YQLi03}. If $N_{b_1}=N_{b_2}=0$, the system degenerates to the
spin-1/2 fermions and is solved by Yang \cite{CNYang67}. If $N_{b_2}=0$, the
system degenerates to the mixture of scalar bosons and spin-1/2 fermions
which is solved by Lai and Yang \cite{CKLai71}. In this paper, we consider
the case that all the particles numbers $N_{b_1}, N_{b_2}, N_{f_1}$ and $%
N_{f_2}$ are not zero. We fist derive the two-body scattering
matrix by using the coordinate Bethe ansatz method and prove the
integrability of the system. Then we determine the Bethe ansatz
equations and the energy spectrum by using the nest quantum
inverse scattering methods.

\subsection{Coordinate Bethe ansatz}

In the framework of coordinate Bethe ansatz, the wave function of the system
described by a set of quasi-momenta $\{k_j\}$ can be written as \cite%
{CNYang67,CKLai71}
\begin{eqnarray}
\Psi(x_1s_1, \cdots,
x_Ns_N)=\sum_{Q,P}\theta(x_{Q_1}<\cdots<x_{Q_N})A_{s_1\cdots s_N} (Q,
P)e^{i\sum_{l=1}^{N}k_{P_l}x_{Q_l}},
\end{eqnarray}
where $Q=(Q_1, \cdots, Q_N)$ and $P=(P_1, \cdots, P_N)$ are the permutations
of the integers $1,\cdots, N$, $N$ is the total number of particles, $N=
N_{b_1}+ N_{b_2}+ N_{f_1} + N_{f_2}$, $\theta(x_{Q_1}<\cdots
<x_{Q_N})=\theta( x_{Q_N}-x_{Q_{N-1}})\cdots \theta( x_{Q_2}-x_{Q_1})$ and $%
\theta(x-y)$ is the step function. The wave function is supersymmetric under
permutating both the coordinates and the spins (or isospins) of two
particles. The wave function is continuous but its derivative jumps when two
atoms touch. With the standard coordinate Bethe ansatz procedure, we obtain
the two-body scattering matrix as
\begin{eqnarray}
S_{jl}(k_j-k_l)=\frac{k_j-k_l-icP^s_{jl}}{k_j-k_l+ic},  \label{s1}
\end{eqnarray}
where $P_{jl}^s$ is the spin super permutation operator with the definition $%
[P^s_{jl}]_{\alpha\mu}^{\beta\nu}=(-1)^{\epsilon_{\alpha}
\epsilon_{\beta}}\delta_{\alpha\nu}\delta_{\mu\beta}$, the $\alpha$ and $\mu$
are the row indices, and $\beta$ and $\nu$ are the column indices. Here $%
\epsilon_{\alpha}$ is the grassmann number, $\epsilon_{a}=0$ for bosons and $%
\epsilon_{a}=1$ for fermions. The scattering matrix satisfies the
super or graded Yang-Baxter equation
\cite{r1,zhao,eks,eks0,eks1,eks2,r2}
\begin{eqnarray}
S_{12}(k_1-k_2) S_{13}(k_1-k_3) S_{23}(k_2-k_3)= S_{23}(k_2-k_3)
S_{13}(k_1-k_3) S_{12}(k_1-k_2). \label{gybe}
\end{eqnarray}
which ensures the integrability of the model (\ref{h1}). The Yang-Baxter
equation (\ref{gybe}) can also be written out explicitly as
\begin{eqnarray}
&&S_{12}(k_1-k_2)_{a_1a_2}^{b_1b_2} S_{13}(k_1-k_3)_{b_1a_3}^{c_1b_3}
S_{23}(k_2-k_3)_{b_2b_3}^{c_2c_3} (-)^{(\epsilon _{b_1}+\epsilon
_{c_1})\epsilon _{b_2}}  \nonumber \\
&&\quad = S_{23}(k_2-k_3)_{a_2a_3}^{b_2b_3}
S_{13}(k_1-k_3)_{a_1b_3}^{b_1c_3}
S_{12}(k_1-k_2)_{b_1b_2}^{c_1c_2}(-)^{(\epsilon _{a_1} +\epsilon
_{b_1})\epsilon _{b_2}}.
\end{eqnarray}

With the periodic boundary conditions of the wave function, we obtain the
following eigenvalue equations
\begin{eqnarray}
&&S_{jN}(k_j-k_N)S_{jN-1}(k_j-k_{N-1})\cdots S_{jj+1}(k_j-k_{j+1}) \nonumber \\
&& \quad\quad \times S_{jj-1}(k_j-k_{j-1})\cdots
S_{j1}(k_j-k_{1})e^{ik_jL}\xi_0=\xi_0,  \label{ei}
\end{eqnarray}
where $\xi_0$ is the amplitude of initial state wave function.

\subsection{Algebraic Bethe ansatz}

Now, we derive the exact solutions of the system by using the
graded nested quantum inverse scattering method
\cite{r1,zhao,eks,eks0,eks1,eks2,r2}. We consider the exact
solutions of the system with BBFF grading, that is the grassmann
parities for the four bases are $\epsilon_1=\epsilon_2=0$ and
$\epsilon_3=\epsilon_4=1$. Please note choosing different bases
is equivalent to choosing different highest weight represents
when deriving the Bethe ansatz equations. The Bethe ansatz
equations with different gradings can change into each others by
using some transformations \cite{eks1}.

The matrix form of the scattering matrix $S_{0j}(\lambda)$ in the
space $0$ is
\begin{eqnarray}
S_{j}(\lambda)=\left(
\begin{array}{cccc}
a(\lambda)-b(\lambda) e_j^{11} & -b(\lambda) e_j^{21} &
-b(\lambda) e_j^{31}
& -b(\lambda) e_j^{41} \\
-b(\lambda) e_j^{12} & a(\lambda)-b(\lambda) e_j^{22} &
-b(\lambda)e_j^{32}
& -b(\lambda)e_j^{42} \\
-b(\lambda)e_j^{13} & -b(\lambda)e_j^{23} & a(\lambda)+
b(\lambda)e_j^{33} &
b(\lambda)e_j^{43} \\
-b(\lambda)e_j^{14} & -b(\lambda)e_j^{24} & b(\lambda)e_j^{34} &
a(\lambda)+
b(\lambda)e_j^{44}%
\end{array}%
\right), \label{scatt}
\end{eqnarray}
where the matrix $e_j^{\alpha \beta }$ acts on the $j$-th space
with its elements defined as $(e_j^{\alpha \beta
})_{\mu\nu}=\delta_{\alpha\mu}\delta_{\beta\nu}$, $
a(\lambda)=\lambda/(\lambda+ic)$ and $b(\lambda)=ic/(\lambda+ic)$.
The quantity (\ref{scatt}) is also called as the Lax operator acting on the $%
j$-th space. Introduced the braid scattering matrix
$R_{12}(\lambda)=P^s_{12}S_{12}(\lambda)$, which satisfies the
braid Yang-Baxter equation,
\begin{eqnarray}
R_{12}(\lambda-u)R_{23}(\lambda)R_{12}(u) =
R_{23}(u)R_{12}(\lambda)R_{23}(\lambda-u).  \label{15}
\end{eqnarray}
We follow the graded nested algebraic Bethe ansatz method to
solve the eigenvalue equation (\ref{ei}). The monodromy matrix is
defined as
\begin{eqnarray}
&&T_N(\lambda)= S_{0j}(\lambda -k_j)S_{0N}(\lambda -k_N)\cdots
S_{0j+1}(\lambda -k_{j+1}) S_{0j-1}(\lambda -k_{j-1})\cdots
S_{01}(\lambda -k_1)  \nonumber \\
&& \qquad\; =\left(
\begin{array}{cccc}
A_{11}(\lambda) & A_{12}(\lambda) & A_{13}(\lambda) & B_1(\lambda) \\
A_{21}(\lambda) & A_{22}(\lambda) & A_{23}(\lambda) & B_2(\lambda) \\
A_{31}(\lambda) & A_{32}(\lambda) & A_{33}(\lambda) & B_3(\lambda) \\
C_{1}(\lambda) & C_{2}(\lambda) & C_{3}(\lambda) & D(\lambda)%
\end{array}
\right),  \label{m1}
\end{eqnarray}
where $0$ means the auxiliary space and $l=1,2,\cdots N$ mean the
quantum spaces. From Eq. (\ref{15}), we can prove that the
monodromy matrix (\ref{m1}) satisfies the Yang-Baxter relation
\begin{eqnarray}
R_{12}(\lambda-u) [T_N(\lambda)\otimes_{s}T_N(u)]
=[T_N(u)\otimes_{s}T_N(\lambda)] R_{12}(\lambda-u),  \label{ybe1}
\end{eqnarray}
where $\otimes_s$ means the super or graded tensor-product as
$[A\otimes_s B]_{ac}^{bd}=(-1)^{(\epsilon _a+\epsilon _b)\epsilon
_c}A_{ab}B_{cd}$, $a$ and $c$ are the row indices, and $b$ and
$d$ are the column indices. Using indices, the Yang-Baxter
relation (\ref{ybe1}) can also be written as
\begin{eqnarray}
&&R_{12}(\lambda-u)_{a_1a_2}^{b_1b_2} T_N(\lambda)_{b_1}^{c_1}
T_N(u)_{b_2}^{c_2}(-1)^{(\epsilon_{b_1}+\epsilon_{c_1})\epsilon_{b_2}}
\nonumber \\
&& \quad\quad
=T_N(u)_{a_1}^{b_1}T_N(\lambda)_{a_2}^{b_2}R_{12}(%
\lambda-u)_{b_1b_2}^{c_1c_2}
(-1)^{(\epsilon_{a_1}+\epsilon_{b_1})\epsilon_{a_2}},
\label{ttybe1}
\end{eqnarray}
where all the repeated indices should be summed. The elements of
scattering matrix $S_{ij}(\lambda)_{a_1a_2}^{b_1b_2}$ are not
zero only with the conditions (1) $a_1=a_2=b_1=b_2$ or (2)
$a_1=b_1, a_2=b_2$ or (3) $a_1=b_2, a_2=b_1$. These properties
will be used in deriving the commutation relations. The transfer
matrix $t(\lambda)$ of the system is defined as the supertrace of
the monodromy matrix (\ref{m1}) in the auxiliary space,
\begin{eqnarray}
t(\lambda)=strT_N(\lambda) =\sum_{a=1}^4 (-1)^{\epsilon
_a}T_N(\lambda)_{a}^{a}=A_{11}(\lambda)+
A_{22}(\lambda)-A_{33}(\lambda)-D_{11}(\lambda).  \label{tran}
\end{eqnarray}
From the Yang-Baxter relation (\ref{ybe1}), we can prove that the
transfer matrices with different spectral parameters commute with
each other $[t (u),t (v)]=0$. Thus the system has infinite
conserved quantities and is integrable. The eigenvalue problem
(\ref{ei}) is therefore reduced to
\begin{eqnarray}
-str_0 T_N(k_j)e^{ik_jL}\xi_0=\xi_0.  \label{2tran}
\end{eqnarray}

We choose the local vacuum state as $|0\rangle_j=(0,0,0,1)^t$
where $t$ means the transpose.
The global vacuum state is constructed as $|0\rangle=\otimes_{j=1}^{N}|0%
\rangle_j$. Acting the monodromy matrix (\ref{m1}) on this vacuum
state, we have
\begin{eqnarray}
T_N(\lambda)|0\rangle =\left(
\begin{array}{cccc}
\prod_{l=1}^{N} a(\lambda-k_l) & 0 & 0 & 0 \\
0 & \prod_{l=1}^{N} a(\lambda-k_l) & 0 & 0 \\
0 & 0 & \prod_{l=1}^{N} a(\lambda-k_l) & 0 \\
C_1(\lambda) & C_2(\lambda) & C_3(\lambda) & 1%
\end{array}%
\right) |0\rangle.  \label{2act}
\end{eqnarray}
We see that the elements $A_{11}(\lambda), A_{22}(\lambda),
A_{33}(\lambda)$ and $D(\lambda)$ acting on this vacuum state
give the eigenvalues. The
elements $B_{a}(\lambda)$ acting on the vacuum state are zero. The elements $%
C_{a}(\lambda)$ acting on the vacuum state give nonzero values
and can be
regarded as the creation operators. We assume the eigenstates of the system (%
\ref{h1}) are obtained by applying the creation operators $C_a$
on the vacuum state as
\begin{equation}
|\lambda_1,\cdots,\lambda_{N_1}|F\rangle =C_{a_1}(\lambda_1)\cdots
C_{a_{N_1}}(\lambda_{N_1})|0\rangle F^{a_{N_1}\cdots a_1},
\label{2st}
\end{equation}
where $F^{a_{N_1} \cdots a_1}$ is a function of the spectral parameters $%
\lambda_j$ and $N_1$ is the number of creation operators. When
the transfer matrix acting on the Bethe states (\ref{2st}), we
need the commutation relations between $A_{11}$, $A_{22}$,
$A_{33}$, $D$ and $C_{a}$. From the Yang-Baxter relation
(\ref{ttybe1}) and using the properties of the $R$ matrix, we
find following commutation relations
\begin{eqnarray}
&&D(u)C_c(\lambda)=\frac {1}{a(\lambda-u)}C_c(\lambda)D(u)- \frac {%
b(\lambda-u)}{a(\lambda-u)}C_c(u)D(\lambda),  \label{2cc1} \\
&&A_{ab}(u)C_c(\lambda)=(-1)^{\epsilon_a \epsilon_e +\epsilon_a +
\epsilon_b} \frac {R^{(1)}_{BBF}(u-\lambda)_{de}^{cb}}{a(u-\lambda)}
C_{e}(\lambda)A_{ad}(u) \nonumber \\
&&\quad\quad \quad \quad \quad \quad \quad  -(-1)^{(\epsilon_a+1)(\epsilon_b+1)} \frac{%
b(u-\lambda)}{a(u-\lambda)}C_b(u)A_{ac}(\lambda),  \label{2cc2} \\
&&C_{a_1}(u)C_{a_2}(\lambda)= R_{FFB}^{(1)}(u-\lambda)_{b_1b_2}^{a_2a_1}
C_{b_2}(\lambda)C_{b_1}(u),  \label{2cc3}
\end{eqnarray}
where all the indices take values 1, 2 and 3. The first nesting $R$ matrices
are defined as
\begin{eqnarray}
R^{(1)}_{FFB}(u)=b(u)+a(u)P_{FFB}^{(1)}, \quad
R^{(1)}_{BBF}(u)=-b(u)+a(u)P_{BBF}^{(1)},
\end{eqnarray}
where ${P^{(1)}_{FFB}}$ and ${P^{(1)}_{BBF}}$ are the $9\times 9$ super
permutation matrices for the grading $\epsilon_1=\epsilon_2=1, \epsilon_3=0$
and $\epsilon_1=\epsilon_2=0, \epsilon_3=1$, respectively.

Acting the transfer matrix (\ref{2tran}) on the assumed eigenstate (\ref{2st}%
), applying repeatedly the commutation relations (\ref{2cc1}) - (\ref{2cc3})
and using the result (\ref{2act}), we have
\begin{eqnarray}
&&t (u)|\lambda_1, \cdots \lambda_{N_1}| F \rangle = \left\{\prod
_{j=1}^{N_1} \frac { 1}{a(u-\lambda_j)} \prod _{l=1}^N a(u-k_l)
t^{(1)}(u) \right. \nonumber \\
&&\quad\quad\left. -\prod _{j=1}^{N_1} \frac { 1}{a(\lambda_j-u)}
\right\} |\lambda_1,\cdots,\lambda_{N_1}|F\rangle +u.t.,
\label{op1}
\end{eqnarray}
where $t^{(1)}(u)$ is the first nesting transfer matrix and
$u.t.$ means the unwanted terms. If the unwanted terms cancel
with each other, which gives following Bethe ansatz equations
\begin{eqnarray}
\prod _{j=1,\neq \alpha}^{N_1} \frac{a(\lambda_{\alpha}-\lambda_j)}{%
a(\lambda_j-\lambda_{\alpha})} \prod _{l=1}^N \frac {1}{a(\lambda_{%
\alpha}-k_l)} F^{b_{N_1}\cdots b_1} = {t^{(1)}(\lambda_{\alpha})}_{a_1
\cdots a_{N_1}}^{b_1 \cdots b_{N_1}}F^{a_{N_1}\cdots a_1}, \quad
\alpha=1,2,\cdots, {N_1},  \label{2bea11}
\end{eqnarray}
then the assumed states (\ref{2st}) are the eigenstates of the
transfer matrix $t(u)$ and the corresponding eigenvalus are given
by the first term in Eq. (\ref{op1}).

Now, seeking the eigenvalues of $t(u)$ becomes seeking the eigenvalues of $%
t^{(1)}(u)$. The elements of first nested transfer matrix
$t^{(1)}(u)$ can be written out explicitly
\begin{eqnarray}
&&t^{(1)}(u, \{\lambda\} )^{b_1\cdots b_{N_1}}_{a_1\cdots a_{N_1}}
=(-1)^{\epsilon_{c_0}} S^{(1)}_{0{N_1}}(u-\lambda_{N_1})_{{c_0}
b_{N_1}}^{c_{{N_1}-1}a_{N_1}} S^{(1)}_{0{N_1}-1}(u-\lambda_{{N_1}-1})_{c_{{%
N_1}-1} b_{{N_1}-1}}^{c_{{N_1}-2}a_{{N_1}-1}} \cdots
S^{(1)}_{01}(u-\lambda_1)_{c_1 b_1}^{{c_0}a_1}  \nonumber \\
&&\quad\quad \times (-1)^{\epsilon _{c_0} \sum_{i=1}^{{N_1}-1}(
\epsilon_{b_i} + 1) + \sum_{i=1}^{{N_1}-1} \epsilon _{c_i}
(\epsilon _{b_i}+1)}.  \label{tcd}
\end{eqnarray}
Here all the indices $c_i$ are summed over and $S^{(1)}(u)={P_{BBF}^{(1)}}%
R_{BBF}^{(1)}(u )$. In order to interpret $%
t^{(1)}(u) $ as the supertrace of the monodromy matrix, we define
a new graded tensor product $ [F \bar{\otimes} G]_{ac}^{bd}=F_a^b
G_c^d (-1)^{(\epsilon_a+\epsilon_b)(\epsilon_c+1)}$. This new
graded tensor-product switches even and odd grassmann parities.
Meanwhile, the first nesting monodromy matrix is defined as
\begin{eqnarray}
&&T_{N_1}^{(1)}(u )=
S^{(1)}_{0{N_1}}(u-\lambda_{N_1})\bar{\otimes}
S^{(1)}_{0{N_1}-1}(u-\lambda_{{N_1}-1})\bar{\otimes} \cdots
\bar{\otimes}
S^{(1)}_{01}(u-\lambda_1)  \nonumber \\
&&\qquad \quad = \left(
\begin{array}{ccc}
A_{11}^{(1)}(u ) & A_{12}^{(1)}(u ) & B_1^{(1)}(u ) \\
A_{21}^{(1)}(u ) & A_{22}^{(1)}(u ) & B_2^{(1)}(u ) \\
C_{1}^{(1)}(u ) & C_{2}^{(1)}(u ) & D^{(1)}(u )%
\end{array}
\right),  \label{2m2}
\end{eqnarray}
which satisfies the graded Yang-Baxter relation,
\begin{eqnarray}
\hat{r}(u-v)
\left[T_{N_1}^{(1)}(u)\bar{\otimes}T_{N_1}^{(1)}(v)\right]
=\left[T_{N_1}^{(1)}(v)\bar{\otimes}T_{N_1}^{(1)}(u)\right]
\hat{r}(u-v), \label{2ybe2}
\end{eqnarray}
where the $\hat{r}$-matrix is
$\hat{r}_{ac}^{bd}=-b(u)\delta_{ab}\delta_{cd} +a(u)
\delta_{ad}\delta_{bc} (-1)^{\epsilon_a +\epsilon_c +\epsilon_a
\epsilon_c}$. Then the transfer matrix (\ref{tcd}) is the
supertrace of the first nesting monodromy matrix (\ref{2m2})
\begin{eqnarray}
t^{(1)}(u, \{\lambda\} )^{b_1\cdots b_{N_1}}_{a_1\cdots a_{N_1}}= str
T^{(1)}_{N_1} (u) =A_{11}^{(1)}(u )+A_{22}^{(1)}(u )- D^{(1)}(u ) .
\label{2tcd}
\end{eqnarray}

We choose $|0\rangle_j^{(1)}=(0,0,1)^t$ as the local reference state for the
first nesting. The global reference state is $|0\rangle^{(1)}=\bar{\otimes}%
_{j=1}^{N_1}|0\rangle_j^{(1)}$. Acting the first nesting monodromy matrix (%
\ref{2m2}) on this reference state, we have
\begin{eqnarray}
T_{N_1}^{(1)}(u)|0\rangle^{(1)} =\left(
\begin{array}{ccc}
\prod_{l=1}^{N_1} a(u-\lambda_l) & 0 & 0 \\
0 & \prod_{l=1}^{N_1} a(u-\lambda_l) & 0 \\
C_1^{(1)}(u) & C_2^{(1)}(u) & 1%
\end{array}%
\right) |0\rangle^{(1)}.  \label{2act2}
\end{eqnarray}
Assume the eigenstates of the first nesting transfer matrix are
\begin{equation}
|\lambda_1^{(1)},\cdots,\lambda_{N_2}^{(1)}|G\rangle
=C_{b_1}^{(1)}(\lambda_1^{(1)})\cdots
C_{b_{N_2}}^{(1)}(\lambda_{N_2}^{(1)})|0\rangle^{(1)} G^{b_{N_2}\cdots b_1},
\label{2st2}
\end{equation}
where ${N_2}$ is the number of creation operators. From the Yang-Baxter
relation (\ref{2ybe2}), we obtain the following commutation relations
\begin{eqnarray}
&&D^{(1)}(u)C^{(1)}_c(\lambda)=\frac {1}{a(\lambda-u)}C^{(1)}_c(%
\lambda)D^{(1)}(u) -\frac {b(\lambda-u)}{a(\lambda-u)}C^{(1)}_c(u)D^{(1)}(%
\lambda), \\
&&A_{ab}^{(1)}(u)C^{(1)}_c(\lambda)= \frac {R^{(2)}_{BB}(u-\lambda)_{de}^{cb}%
}{a^{(1)}(u-\lambda)}C_{e}^{(1)}(\lambda)A_{ab_1}^{(1)}(u) +\frac{%
b(u-\lambda)}{a(u-\lambda)}C_b^{(1)}(u)A_{ac}^{(1)}(\lambda), \\
&&C_{b_1}^{(1)}(u)C_{b_2}^{(1)}(\lambda)=
R^{(2)}_{FF}(u-\lambda)_{c_1c_2}^{b_2b_1}C_{c_2}^{(1)}(\lambda)
C_{c_1}^{(1)}(u).
\end{eqnarray}
Here, all the indices take values 1 and 2. The second nesting $R$
matrices are $R^{(2)}_{FF}(u)=b(u)+a(u)P_{FF}^{(2)}$ and
$R^{(2)}_{BB}(u)=-b(u)+a(u)P_{BB}^{(2)}$, where ${P^{(2)}_{FF}}$
and ${P^{(2)}_{BB}}$ are the $4\times 4$ super
permutation matrices for the grading $\epsilon_1=\epsilon_2=1$ and $%
\epsilon_1=\epsilon_2=0$, respectively.
$[P^{(2)}_{FF}]_{ac}^{bd}=-\delta_{ad}\delta_{bc}$,
$[P^{(2)}_{BB}]_{ac}^{bd}=\delta_{ad}\delta_{bc}$. Acting the
first nesting transfer matrix (\ref{2tcd}) on the assumed states
(\ref{2st2}), we have
\begin{eqnarray}
&&t^{(1)}(u)|\lambda_1^{(1)}\cdots \lambda_{N_2}^{(1)}|G\rangle =
\left\{\prod _{j=1}^{N_2} \frac {1}{a(u-\lambda_j^{(1)})} \prod
_{l=1}^{N_1}
a(u-\lambda_l) t^{(2)}(u)\right. \nonumber \\
&&\quad\quad\quad \left. -\prod _{j=1}^{N_2} \frac { 1}{a(\lambda_j^{(1)}-u)%
} \right\} |\lambda_1^{(1)},\cdots,\lambda_{N_2}^{(1)}|G\rangle +u.t.,
\label{op2}
\end{eqnarray}
where $t^{(2)}(u )$ is the second nesting transfer matrix
\begin{eqnarray}
t^{(2)}(u )=str S_{0{N_2}}^{(2)}(u-\lambda_{N_2}^{(1)})S_{0{N_2}%
-1}^{(2)}(u-\lambda_{{N_2}-1}^{(1)})\cdots S_{01}^{(2)}(u-\lambda_1^{(1)}).
\end{eqnarray}
If the unwanted terms cancel with each other, which give following Bethe
ansatz equations
\begin{eqnarray}
\prod _{j=1,\neq \alpha}^{N_2} \frac {a(\lambda_{\alpha}^{(1)}-%
\lambda_j^{(1)})} {a(\lambda_j^{(1)}-\lambda_{\alpha}^{(1)})} \prod
_{l=1}^{N_1} \frac {1}{a(\lambda_{\alpha}^{(1)}-\lambda_l)} G^{b_{N_2}\cdots
b_1} ={t^{(2)}(\lambda_{\alpha}^{(1)})}_{a_1 \cdots a_{N_2}}^{b_1 \cdots
b_{N_2}}G^{a_{N_2}\cdots a_1}, \quad \alpha=1,2,\cdots, {N_2},
\label{2bea22}
\end{eqnarray}
the assumed states (\ref{2st2}) are the eigenstates of the first nesting
transfer matrix (\ref{2tcd}). Now, the remanent problem is finding the
eigenvalues of second nesting transfer matrix $t^{(2)}(u )$.

The second nesting scattering matrix is $S^{(2)}(u)= a(u)-b(u)P^{(2)}_{BB}$.
The second nesting monodromy matrix is
\begin{eqnarray}
T_{N_2}^{(2)}(u )=
S_{0{N_2}}^{(2)}(u-\lambda_{N_2}^{(1)})S_{0{N_2}
-1}^{(2)}(u-\lambda_{{N_2}-1}^{(1)})\cdots
S_{01}^{(2)}(u-\lambda_1^{(1)})
 =\left(
\begin{array}{cc}
A^{(2)}(u ) & B^{(2)}(u ) \\
C^{(2)}(u ) & D^{(2)}(u )%
\end{array}
\right),  \label{2m3}
\end{eqnarray}
which satisfies the graded Yang-Baxter relations
\begin{eqnarray}
R_{12}^{(2)}(u-v) [T_{N_2}^{(2)}(u)\bar{\otimes} T_{N_2}^{(2)}(v)]
=[T_{N_2}^{(2)}(v)\bar{\otimes} T_{N_2}^{(2)}(u)] R_{12}^{(2)}(u-v),
\label{2ybe3}
\end{eqnarray}
where $R^{(2)}(u)=P^{(2)}_{BB}S^{(2)}(u)$ is the second nesting braid $R$
matrix. Here the graded tensor-product degenerates to the ordinal
tensor-product because that only the bosonic gradings are left. The second
nesting transfer matrix is the supertrace of the corresponding monodromy
matrix
\begin{eqnarray}
t^{(2)}(u )=str T_{N_2}^{(2)}(u)=A^{(2)}(u )+D^{(2)}(u ).
\end{eqnarray}
We choose $|0\rangle_j^{(2)}=(0,1)^t$ as the local vacuum state for the
second nesting. The global vacuum state is $|0\rangle^{(2)}=%
\otimes_{j=1}^{N_2}|0\rangle_j^{(2)}$. Acting the second nesting momodromy
matrix (\ref{2m3}) on this vacuum state, we have
\begin{eqnarray}
T_{N_2}^{(2)}(u)|0\rangle^{(2)} =\left(
\begin{array}{cc}
\prod_{l=1}^{N_2} a^{(1)}(u-\lambda_l^{(1)}) & 0 \\
C^{(2)}(u) & \prod_{l=1}^{N_2} [a^{(1)}(u-\lambda_l^{(1)}) -
b^{(1)}(u-\lambda_l^{(1)})]%
\end{array}%
\right) |0\rangle^{(2)}.  \label{2act3}
\end{eqnarray}
Assume the eigenstates of the second nesting transfer matrix $t^{(2)}(u )$
are
\begin{equation}
|\lambda_1^{(2)},\cdots,\lambda_{N_3}^{(2)}\rangle
=C^{(2)}(\lambda_1^{(2)})\cdots C^{(2)}(\lambda_{N_3}^{(2)})|0\rangle^{(2)},
\label{2st3}
\end{equation}
where $N_3$ is the number of creation operators. From the Yang-Baxter
relation (\ref{2ybe3}), we obtain following commutation relations
\begin{eqnarray}
&&D^{(2)}(u)C^{(2)}(\lambda)=\frac {a(\lambda-u)-b(\lambda-u)}{a(\lambda-u)}%
C^{(2)}(\lambda)D^{(2)}(u) +\frac {b(\lambda-u)}{a(\lambda-u)}%
C^{(2)}(u)D^{(2)}(\lambda),  \label{4cc1} \\
&&A^{(2)}(u)C^{(2)}(\lambda)=\frac
{a(u-\lambda)-b(u-\lambda)}{a(u-\lambda)}
C^{(2)}(\lambda)A^{(2)}(u) +\frac{b(u-\lambda)}{a(u-\lambda)}%
C^{(2)}(u)A^{(2)}(\lambda),  \label{4cc2} \\
&&C^{(2)}(u)C^{(2)}(\lambda)= C^{(2)}(\lambda) C^{(2)}(u).
\label{4cc3}
\end{eqnarray}
The second nesting transfer matrix $t^{(2)}(u )$ acting the assumed states (%
\ref{2st3}) gives
\begin{eqnarray}
t^{(2)}(u)C^{(2)}(\lambda_1^{(2)})\cdots
C^{(2)}(\lambda_{N_3}^{(2)})|0\rangle^{(2)}   = \left\{\prod _{j=1}^{N_3} \frac {a(u-\lambda_j^{(2)})-b(u-%
\lambda_j^{(2)})}{a(u-\lambda_j^{(2)})} \prod _{l=1}^{N_2}
a(u-\lambda_l^{(1)}) \right.  \nonumber \\
 \left. +\prod _{j=1}^{N_3} \frac {a(\lambda_j^{(2)}-u)
-b(\lambda_j^{(2)}-u)}{a(\lambda_j^{(2)}-u)} \prod _{l=1}^{N_2} \left[%
a(u-\lambda_l^{(1)}) -b(u-\lambda_l^{(1)})\right]\right\}
|\lambda_1^{(2)},\cdots,\lambda_{N_3}^{(2)}\rangle +u.t..  \label{op3}
\end{eqnarray}
If the unwanted terms cancel, the assumed states (\ref{2st3}) are the
eigenstates of the second nesting transfer matrix $t^{(2)}(u )$, which gives
following Bethe ansatz relations
\begin{eqnarray}
\prod _{j=1,\neq \alpha}^{N_3} \frac {a(\lambda_{\alpha}^{(2)}-%
\lambda_j^{(2)})- b(\lambda_{\alpha}^{(2)}-\lambda_j^{(2)})} {%
a(\lambda_j^{(2)}-\lambda_{\alpha}^{(2)})-
b(\lambda_j^{(2)}-\lambda_{\alpha}^{(2)})} \frac {a(\lambda_j^{(2)}-%
\lambda_{\alpha}^{(2)})} {a(\lambda_{\alpha}^{(2)}-\lambda_j^{(2)})}
= \prod _{l=1}^{N_2} \frac {a(\lambda_{\alpha}^{(2)}-%
\lambda_l^{(1)})- b(\lambda_{\alpha}^{(2)}-\lambda_l^{(1)})} {%
a(\lambda_{\alpha}^{(2)}-\lambda_l^{(1)})}, \label{2bea33}
\end{eqnarray}
where $\alpha=1,2,\cdots, {N_3}.$ Now, the eigenvalues of the
transfer matrix $t(u)$ can be calculated
directly by synthetically considering Eqs. (\ref{op1}), (\ref{op2}) and (\ref%
{op3}).

The forth set of Bethe ansatz equations are obtained from the
eigen-equation (\ref{ei}) as
\begin{eqnarray}
e^{-ik_jL}= \prod_{\alpha=1}^{N_1} a(\lambda_{\alpha}-k_j), \quad
j=1,\cdots, N,  \label{2bea44}
\end{eqnarray}

Put $\lambda_j \rightarrow \lambda_j -ic/2$, $\lambda_j^{(1)} \rightarrow
\lambda_j^{(1)} - i c$ and $\lambda_j^{(2)} \rightarrow \lambda_j^{(2)} -i
c/2$, then the Bethe ansatz equations (\ref{2bea11}), (\ref{2bea22}), (\ref%
{2bea33}) and (\ref{2bea44}) can be written out explicitly as
\begin{eqnarray}
&&e^{ik_jL}= \prod_{\alpha=1}^{N_1} \frac{k_j-\lambda_{\alpha}+\frac{i}{2}c%
} {k_j-\lambda_{\alpha}-\frac{i}{2}c}, \quad j=1,\cdots, N,
\label{2bea1}\\
&&\prod_{l=1}^{N} \frac{\lambda_{\alpha}-k_l-\frac{i}{2}c} {%
\lambda_{\alpha}-k_l+\frac{i}{2}c}=\prod_{\beta=1, \neq \alpha}^{N_1} \frac{%
\lambda_{\alpha}-\lambda_{\beta}-ic }{\lambda_{\alpha}-\lambda_{\beta}+ic }
\prod_{\gamma=1}^{N_2} \frac{\lambda_{\gamma}^{(1)}-\lambda_{\alpha}-\frac{i%
}{2}c} {\lambda_{\gamma}^{(1)}-\lambda_{\alpha}+\frac{i}{2}c }, \quad
\alpha=1,\cdots, {N_1},  \label{2bea2} \\
&&\prod_{\gamma=1}^{N_1} \frac{\lambda_{\beta}^{(1)} -\lambda_{\gamma}-\frac{%
i}{2}c }{\lambda_{\beta}^{(1)} -\lambda_{\gamma}+\frac{i}{2}c} =
\prod_{\rho=1}^{N_3} \frac{\lambda_{\rho}^{(2)}-\lambda_{\beta}^{(1)}+\frac{i%
}{2}c} {\lambda_{\rho}^{(2)}-\lambda_{\beta}^{(1)}-\frac{i}{2}c }, \quad
\beta=1,\cdots, {N_2},  \label{2bea3} \\
&&\prod_{\xi=1}^{N_2} \frac{\lambda_{\gamma}^{(2)} -\lambda_{\xi}^{(1)}+%
\frac{i}{2}c }{\lambda_{\gamma}^{(2)} -\lambda_{\xi}^{(1)}-\frac{i}{2}c}=
\prod_{\eta=1, \neq \gamma}^{N_3} \frac{\lambda_{\eta}^{(2)}-\lambda_{%
\gamma}^{(2)}-ic} {\lambda_{\eta}^{(2)}-\lambda_{\gamma}^{(2)}+ic }, \quad
\gamma=1,\cdots, {N_3},  \label{2bea4}
\end{eqnarray}
where $N=N_{b_1}+N_{b_2}+N_{f_1}+N_{f_2}, N_1=N_{b_1}+N_{b_2}+N_{f_1},
N_2=N_{b_1}+N_{b_2}, N_3=N_{b_1}$. Taking the logarithm of Eqs. (\ref%
{2bea1}) - (\ref{2bea4}), we arrive at
\begin{eqnarray}
&& k_jL = 2\pi I_j - \sum_{b=1}^{N_1} \Xi_{1/2}(k_j-\lambda_b), \quad
j=1,\cdots,N,  \label{2bae10} \\
&& 2\pi J_{a}= \sum_{l=1}^{N} \Xi_{1/2}(\lambda_a-k_l)
-\sum_{b=1}^{N_1} \Xi_{1}(\lambda_a-\lambda_b) +\sum_{c=1}^{N_2}
\Xi_{1/2}(\lambda_a-\lambda_c^{(1)}), \; a=1,\cdots, {N_1},
\label{2bae20} \\
&&2\pi J_{a_1}^{(1)}= \sum_{b=1}^{N_1}
\Xi_{1/2}(\lambda_{a_1}^{(1)}-\lambda_b) -\sum_{c=1}^{N_3}
\Xi_{1/2}(\lambda_{a_1}^{(1)}-\lambda_c^{(2)}), \quad a_1=1,\cdots, {N_2},
\label{2bae30} \\
&& 2\pi J_{a_2}^{(2)}=- \sum_{b=1}^{N_2}
\Xi_{1/2}(\lambda_{a_2}^{(2)}-\lambda_b^{(1)}) +\sum_{c=1}^{N_3}
\Xi_{1}(\lambda_{a_2}^{(2)}-\lambda_c^{(2)}), \quad a_2=1,\cdots, {N_3},
\label{2bae40}
\end{eqnarray}
where $\Xi_{m}(x)=2\tan^{-1}[x/(m c)]$, $I_j$, $J_{a}$,
$J_{a_1}^{(1)}$ and $J_{a_2}^{(2)}$ are integer or half-odd
quantum numbers. Here we have used the formula $\ln
[(k+ic)/(k-ic)]=i[\pi-2 \tan^{-1}(k/c)]$. These equations are the
special case of that in the Ref. \cite{zhou}. If we set
$N_{b_1}=N_{b_2}=0$, the system (1) degenerates to the spin-1/2
fermions model with $\delta$-function potentials and our Bethe
ansatz equations (\ref{2bea1}) - (\ref{2bea4}) are the same as
the ones obtained by Yang \cite{CNYang67}. If we set $N_{b_2}=0$,
the system
(1) degenerates to the bose-fermi mixture considered by Lai and Yang \cite%
{CKLai71}, and our Bethe ansatz equations are the same as their
results.

The eigenvalues of the Hamiltonian (1) are $E=\sum_{j=1}^N k_j^2$,
where possible values of the momentum $k_j$ are determined by the
Bethe ansatz equations (\ref{2bae10}) - (\ref{2bae40}).

\subsection{Corresponding lattice model}

From the transfer matrix $t(u)$, we can construct the
corresponding lattice model by using the standard integrable
theory in the statistic physics. The Hamiltonian can be obtained
by taking the derivative of the logarithmic form of the transfer
matrix at the zero spectral parameter point \cite{cao1}. After
some calculations, we find that the $ \partial \ln t
(u)/(\partial u)|_{u=0}$ gives
the Hamiltonian studied by Essler, Korepin and Schoutens \cite%
{eks,eks0,eks1,eks2} up to a constant. The system (\ref{h1}) is
the continue case of Essler-Korepin-Schoutens (EKS) model. The two-body scattering matrix (\ref%
{scatt}) of present model has the same structure as that of EKS
model. In such a sense, the spin dynamics of present model keeps
some similarities to that of EKS model.

We note that the integrable $SU(n|m)$-supersymmetric quantum spin
chain has been studied by Kulish \cite{r1}. The corresponding
Hamiltonian can be obtained directly from the transfer matrix. If
the charge rapidities tend to zero, our results degenerate into
that obtained by Kulish. The thermodynamics and low-energy limit
of the $SU(n|m)$-invariant spin chain are studied by Saleur
\cite{r2}. These results are also valid for the present model in
the spin sector.

\section{Ground state of the system}

We first consider the case of $c > 0$, which means the interactions among
the particles are repulsive. We use the BBFF grading as a demonstration.
Above we have confined the particles in a finite 1D box with the length $L$.
Some useful properties can be obtained by the analysis of the poles or zeros
of the Bethe ansatz equations in the thermodynamic limit, that is the system
size $L$, particles numbers $N, N_1, N_2$ and $N_3$ tend to infinity, but
the ratios $N/L, N_1/L, N_2/L$ and $N_3/L$ keep finite. For example, if some
$k_j$ are in the upper complex plane, then the left hand side of Eq. (\ref%
{2bea1}) tends to zero when the system size tends to infinity. Thus the
right hand side of Eq. (\ref{2bea1}) should go to zero too. From further
analysis of the Bethe ansatz Eqs. (\ref{2bea1}) - (\ref{2bea4}), we find
that the momentum $k_j$, rapidities $\lambda $ and $\lambda^{(1)}$ are real
at the ground state. The particle number $N_{b_2}=0$ thus the corresponding
rapidity $\lambda^{(2)}$ is zero too. That is to say only one species of
bosons are left or the bosons are totally polarized at the ground state. The
ground state for the bosons is a ferromagnetic state. Meanwhile, the
particle numbers of two species of fermions are equal, $N_{f_1}=N_{f_2}$ and
they form the spin singlet states. The ground state for the fermions is a
antiferromagnetic state. Therefore, the ground state of the system (1) is
partial polarized.

The momentum $k_{j}$, rapidities $\lambda $ and $\lambda ^{(1)}$ at the
ground state satisfy the following coupled equations
\begin{eqnarray}
&&k_{j}L=2\pi I_{j}-\sum_{b=1}^{N_{1}}\Xi _{1/2}(k_{j}-\lambda _{b}),
\label{12bae10} \\
&&2\pi J_{a}=\sum_{l=1}^{N}\Xi _{1/2}(\lambda
_{a}-k_{l})-\sum_{b=1}^{N_{1}}\Xi _{1}(\lambda _{a}-\lambda
_{b})+\sum_{c=1}^{N_{2}}\Xi _{1/2}(\lambda _{a}-\lambda _{c}^{(1)}),
\label{12bae20} \\
&&2\pi J_{a_{1}}^{(1)}=\sum_{b=1}^{N_{1}}\Xi _{1/2}(\lambda
_{a_{1}}^{(1)}-\lambda _{b}).  \label{12bae30}
\end{eqnarray}%
The quantum numbers $I_{j}$ take integer (half-odd integer)
values if $N_{b_{1}}+N_{f_{1}}$ is even (odd), $J_{a}$ take
integer (half-odd integer) values if $N_{b_{1}}+N_{f_{2}}+1$ is
even (odd) and $J_{a_{1}}^{(1)}$ take integer (half-odd integer)
values if $N_{b_{1}}+N_{f_{1}}$ is even (odd). Then quantum
number configuration of the ground state are $\{I_{j}\}
=\left\{-(N-1)/2, -(N-3)/2,\dots, (N-1)/2 \right\}$, $\{J_{a}\}
=\left\{-(N_{1}-1)/2, -(N_{1}-3)/2, \dots, \right.$
$\left.(N_{1}-1)/2 \right\}$ and $\{J_{a_{1}}^{(1)}\}
=\left\{-(N_{2}-1)/2, -(N_{2}-3)/2, \dots, (N_{2}-1)/2\right\}$,
which are symmetrically centered around the origin if
$N_{f_{1}}=N_{f_{2}}$ is odd and $N_{b}=N_{b_{1}}$ is even.

In the thermodynamic limit, the summations become integrations.
The quantum number $I_{j},J_{\alpha }$ and $J_{a_{1}}^{(1)}$
become continue functions of the spectral parameters $k$, $\lambda
$ and $\lambda ^{(1)}$, respectively. Denote the densities of
momentum $k$, rapidities $\lambda $ and $\lambda ^{(1)}$ by $\rho
(k)$, $\rho _{1}(\lambda )$ and $\rho _{1}^{(1)}(\lambda ^{(1)})$,
respectively. Then we have $ \rho (k)=d I_{j}/(L dk)$, $\rho
_{1}(\lambda )=d J_{a}/(L d\lambda )$ and $\rho ^{(1)}(\lambda
^{(1)})= d J_{a_{1}}^{(1)}/ (Ld\lambda ^{(1)})$. Taking the
derivative of Eqs. (\ref{12bae10}) - (\ref{12bae30}), we obtain
the densities of states at the ground state as
\begin{eqnarray}
&&\rho (k)=\frac{1}{2\pi }+\frac{1}{\pi }\int_{-B}^{B}\frac{2c\rho
_{1}(\lambda )d\lambda }{c^{2}+4(k-\lambda )^{2}},  \nonumber \\
&&\rho _{1}(\lambda )=\frac{1}{\pi }\int_{-Q}^{Q}\frac{2c\rho (k)dk}{%
c^{2}+4(\lambda -k)^{2}}-\frac{1}{\pi }\int_{-B}^{B}\frac{c\rho _{1}(\lambda
^{\prime })d\lambda ^{\prime }}{c^{2}+(\lambda -\lambda ^{\prime })^{2}}+%
\frac{1}{\pi }\int_{-D}^{D}\frac{2c\rho ^{(1)}(\lambda ^{(1)})d\lambda ^{(1)}%
}{c^{2}+4(\lambda -\lambda ^{(1)})^{2}},  \nonumber \\
&&\rho ^{(1)}(\lambda ^{(1)})=\frac{1}{\pi }\int_{-B}^{B}\frac{2c\rho
_{1}(\lambda )d\lambda }{c^{2}+4(\lambda ^{(1)}-\lambda )^{2}}.
\end{eqnarray}%
The integral limits $Q,B$ and $D$ are determined by
\[
N/L=\int_{-Q}^{Q}\rho (k)dk,\quad {N_{1}}/L=\int_{-B}^{B}\rho _{1}(\lambda
)d\lambda ,\quad {N_{2}}/L=\int_{-D}^{D}\rho ^{(1)}(\lambda ^{(1)})d\lambda
^{(1)}.
\]%
The quantity $Q$ is the Fermi surface of the system. The densities of energy
and momentum at the ground state are
\[
E/L=\int_{-Q}^{Q}k^{2}\rho (k)dk,\quad P/L=\int_{-Q}^{Q}k\rho (k)dk.
\]%
The magnetization of the fermions $S_{f}^{z}$ and the magnetization of
bosons $\mathcal{T}^{z}$ are
\[
S_{f}^{z}=N_{f_{1}}-N_{f_{2}}=0,\quad \mathcal{T}%
_{b}^{z}=N_{b_{1}}-N_{b_{2}}=N_{b_{1}}.
\]

In the case of $c<0$, besides real solutions, Eqs. (\ref{2bea1}) - (\ref%
{2bea4}) also have complex solutions which are usually called as string
solutions. After some algebraic calculations, we find that at the ground
state, the momentum $k_j$ have the following $2$-string and $N_{b_1}$-string
solutions,
\begin{eqnarray}
&& k_j = \Lambda_j + \frac{ic}{2}(3-2j) + o(e^{-\delta L}),\quad j=1, 2,
\label{string1} \\
&&k_l= \frac{ic}{2}(N_{b_1}+1-2l)+ o(e^{-\delta^\prime L}), \quad l=1, 2
\cdots, N_{b_1},  \label{string2}
\end{eqnarray}
where $\Lambda_j$ is a real parameter, $\delta$ and $\delta^\prime$ are some
positive constants. Eqs. (\ref{string1}) and (\ref{string2}) means that the
fermions form the spin singlet states and the bosons condensed at the zero
momentum point.

\section{Low-lying excitation}

In this section, we consider the low-lying excitations in the
system. We will follow the methods proposed by Takahashi
\cite{takahashi,tak18,tak19}, Essler and Korepin \cite{eks2} very
closely. The low-lying excitations are very rich due to the
complicated solutions of the Bethe ansatz equations. Let us
consider them one by one.

\subsection{Charge-hole excitation}

\begin{figure}[th]
\begin{center}
\includegraphics[height=5.2cm,width=6.3cm]{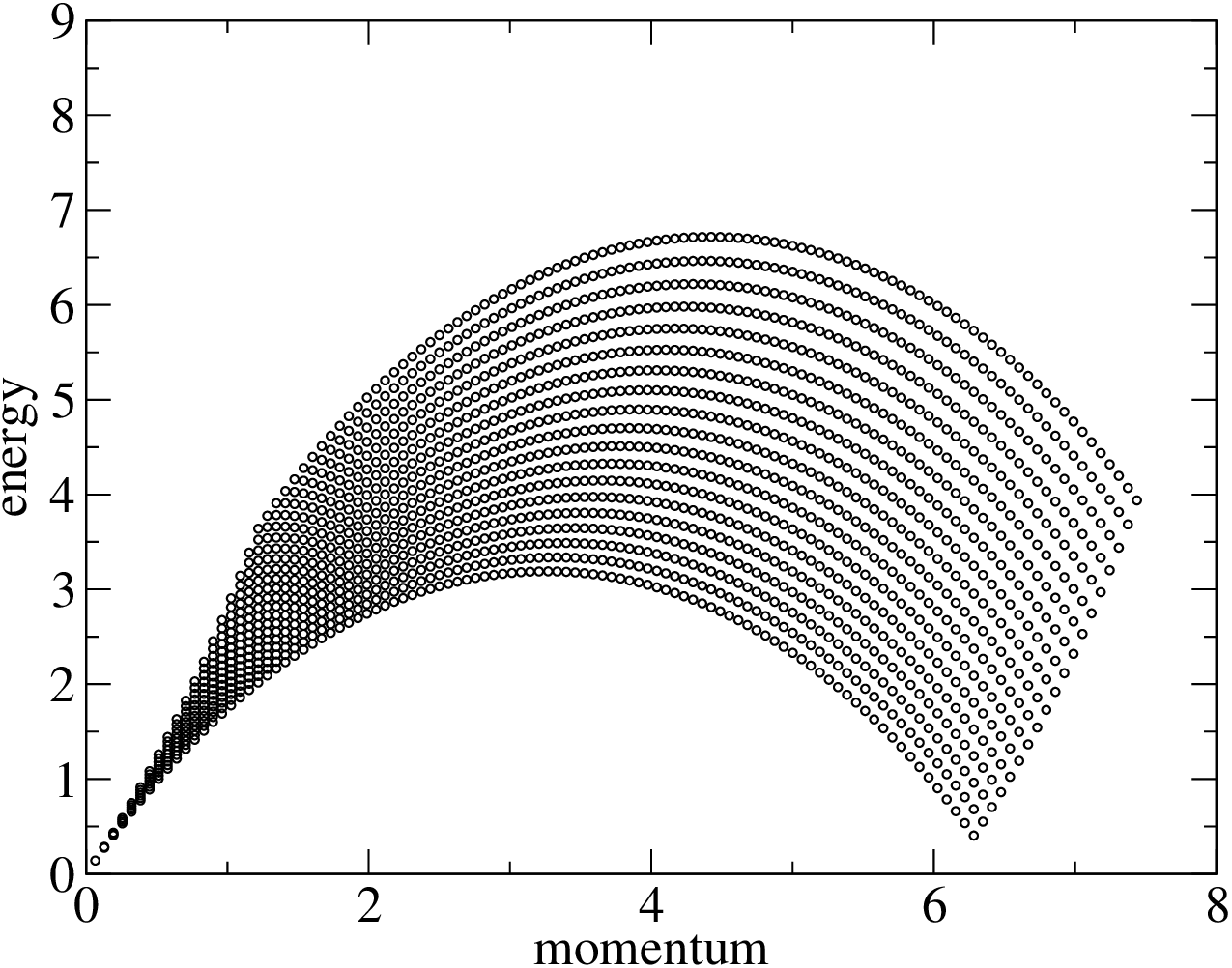}\quad\quad
\includegraphics[height=5.2cm,width=6.3cm]{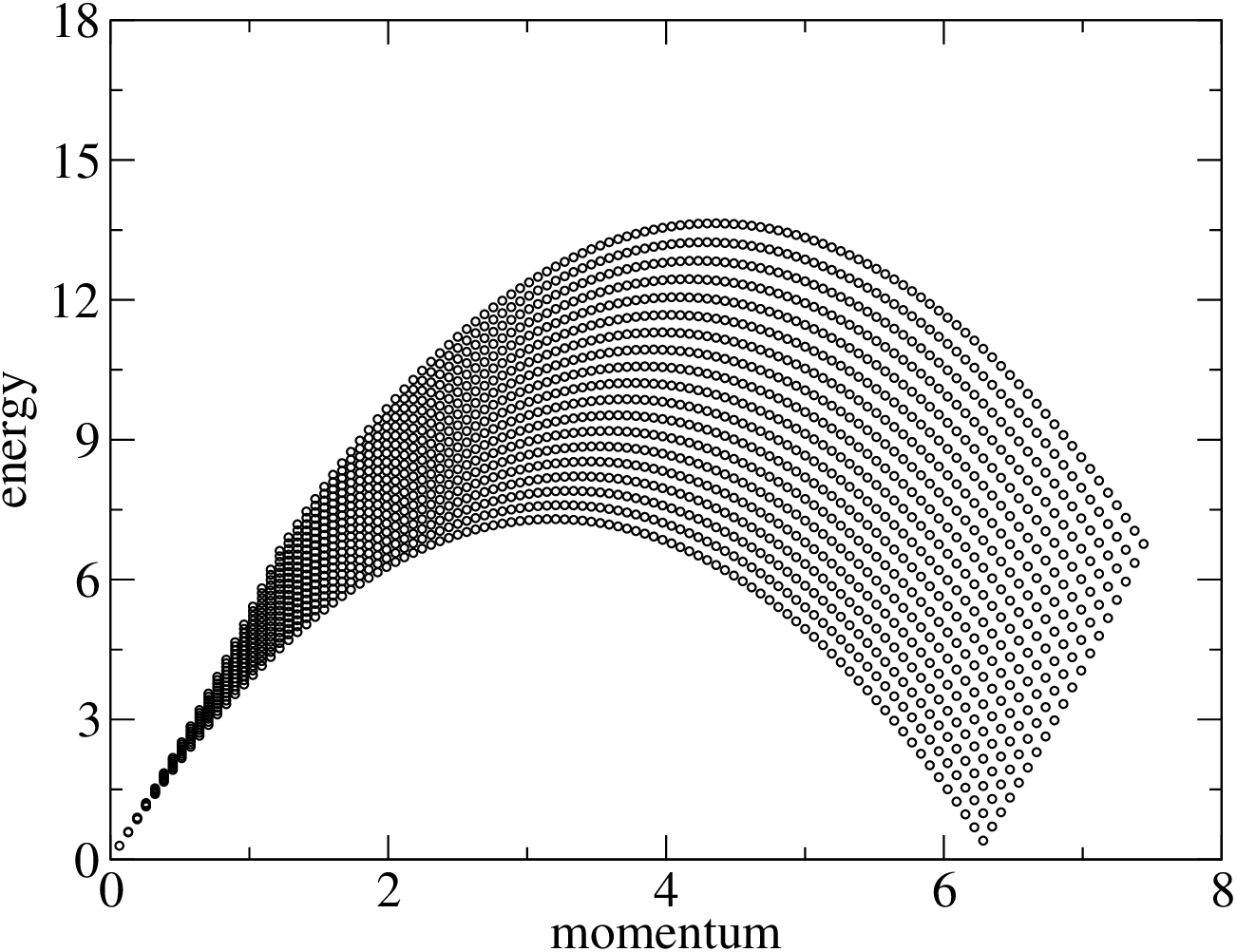}
\end{center}
\caption{The particle-hole excitation spectra calculated for $%
L=98,N_{f_{1}}=N_{f_{2}}=25,N_{b_{1}}=48$, and the coupling
$c=1.0$ (LEFT) and $c=10$ (RIGHT).} \label{fig1}
\end{figure}

\begin{figure}[th]
\begin{center}
\includegraphics[height=5.2cm,width=6.3cm]{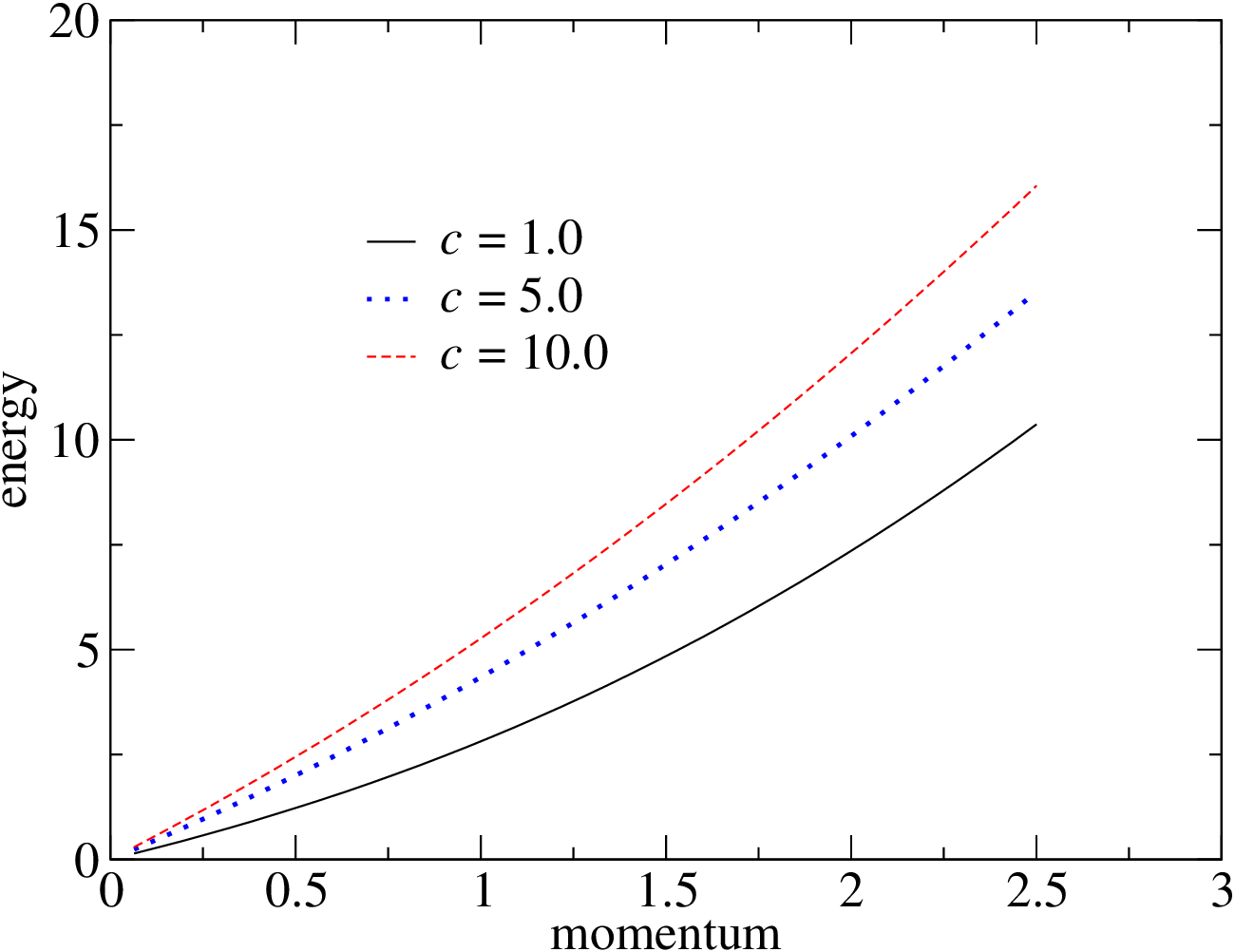}\quad\quad
\includegraphics[height=5.2cm,width=6.3cm]{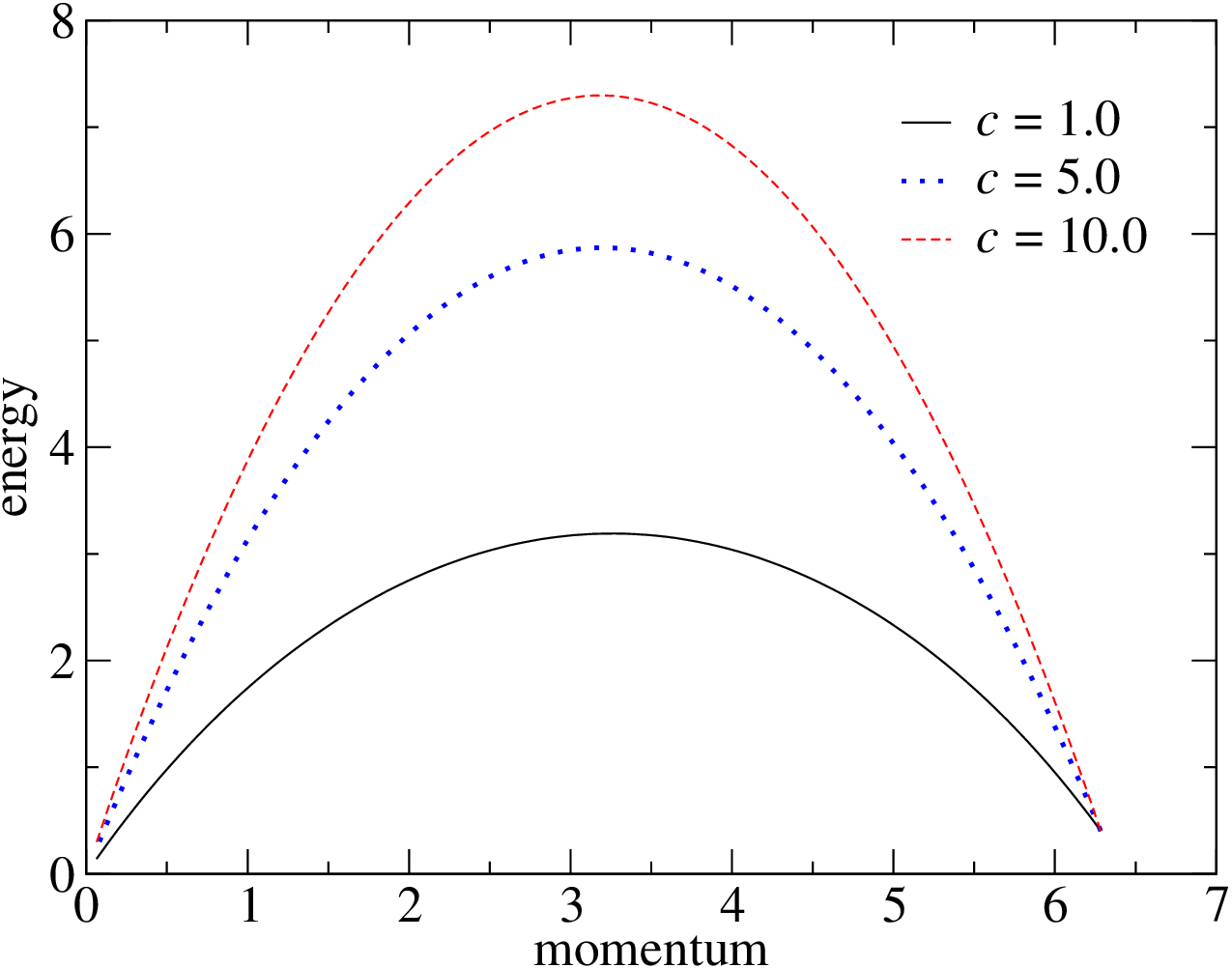}
\end{center}
\caption{The dispersion relations of the elementary excitation:
charge (LEFT) and hole (RIGHT).} \label{fig2}
\end{figure}

The simplest excitation is obtained by removing a quantum number
$I_{j}$ from the sequence $\{I_{j}\}$ and putting it outside the
sequence, i.e.,
\[
\{I_{j}\}=\left\{-\frac{N-1}{2}, \cdots, m-1, m+1, \cdots,
\frac{N-1}{2}, I_{n}\right\},
\]%
where $I_{n}=(N-1)/{2}+n$, and keep the other two quantum number sequence $%
\{J_{a},J_{a_{1}}^{(1)}\}$ unchanged. We call this excitation the
charge-hole excitation since a \textquotedblleft hole" is created under the
Fermi surface and a particle outside the surface. The charge-hole excitation
spectra are shown in Fig. \ref{fig1}. The dispersion relations of charge and
hole are shown in Fig. \ref{fig2}.

In the thermodynamic limit, the densities of states at this excited state
read
\begin{eqnarray}
&&\rho(k) =\frac{1}{2\pi} +\frac{1}{\pi} \int_{-B}^{B} \frac{2c\rho_{1}
(\lambda)d \lambda }{c^2+4(k-\lambda)^2} -\frac{1}{L}\delta(k-k_h),
\nonumber \\
&&\rho_1(\lambda)= \frac{1}{\pi} \int_{-Q}^{Q} \frac{2c\rho (k)d k }{%
c^2+4(\lambda-k)^2}- \frac{1}{\pi} \int_{-B}^{B} \frac{c\rho_{1}
(\lambda^{\prime})d \lambda^{\prime} }{c^2+(\lambda-\lambda^{\prime})^2}
\nonumber \\
&& \quad\quad\quad+\frac{1}{\pi} \int_{-D}^{D} \frac{2c\rho^{(1)}(
\lambda^{(1)} )d \lambda^{(1)} }{c^2+4(\lambda-\lambda^{(1)})^2}+\frac{1}{%
\pi L}\frac{2c}{c^2+4(\lambda-k_p)^2},  \nonumber \\
&&\rho^{(1)}(\lambda^{(1)}) =\frac{1}{\pi} \int_{-B}^{B} \frac{2c\rho_{1}
(\lambda)d \lambda }{c^2+4(\lambda^{(1)}-\lambda)^2},  \label{ex1}
\end{eqnarray}
where $k_h$ is the momentum of the hole and $k_p$ represent the
momentum of quasi-particles. We use the same notations $Q, D$ and
$B$ to present the new integral limits.

Now, we calculate the excitation energy $E_{ex}=E-E_{GS}$, where $E_{GS}$ is
the ground state energy. Follow the methods proposed by Takahashi \cite%
{takahashi,tak18,tak19}, E$\beta$ler and Korepin \cite{eks2}, we define the
differences of the densities of states between the ground state and the
excited state as
\begin{eqnarray}
\sigma_1(k)=L[\rho(k)-\rho_{GS}(k)],
\sigma_2(\lambda)=L[\rho_1(\lambda)-\rho_{1,GS}(\lambda)],
\sigma_3(\lambda^{(1)})=L[\rho^{(1)}(\lambda^{(1)})
-\rho_{GS}^{(1)}(\lambda^{(1)})],
\end{eqnarray}
where $\rho_{GS}$, $\rho_{1,GS}$ and $\rho_{GS}^{(1)}$ are the corresponding
densities at the ground state. The corrections to the densities are
\begin{eqnarray}
\varphi_1=-\delta(k-k_h), \quad \varphi_2=\frac{1}{\pi} \frac{2c}{%
c^2+4(\lambda-k_p)^2}, \quad \varphi_3=0.
\end{eqnarray}
Define the bare energies as
\begin{eqnarray}
\left(
\begin{array}{c}
\varepsilon_1 \\
\varepsilon_2 \\
\varepsilon_3%
\end{array}%
\right)= \left(
\begin{array}{ccc}
1 & -\hat{a}_{1}^{B} & 0 \\
-\hat{a}_{1}^{Q} & 1+ \hat{a}_{2}^{B} & -\hat{a}_{1}^{Q} \\
0 & -\hat{a}_{1}^{B} & 1%
\end{array}%
\right)^{-1} \left(
\begin{array}{c}
k^2 \\
0 \\
0%
\end{array}%
\right),
\end{eqnarray}
where the integral operator ${\hat a}_n^A(x)$ satisfies
\begin{eqnarray}
{\hat a}_n^A(x)=\frac{1}{\pi}\frac{2nc}{n^2c^2+4x^2}, \quad
{\hat a}_n^A(x)*f=\int_{-A}^{A} \frac{1}{\pi}\frac{2nc}{n^2c^2+4(x-y)^2}%
f(y)dy.
\end{eqnarray}
We obtain the excitation energy as
\begin{eqnarray}
E_{ex}=\sum_{\alpha=1}^{3}\int_{-a_{\alpha}}^{a_{\alpha}}
\varepsilon_{\alpha}(\mu)\varphi_{\alpha}d\mu =\epsilon(k_p)-\epsilon(k_h),
\end{eqnarray}
where $\epsilon(k)$ is the dressed energy
\begin{eqnarray}
&&\epsilon(k) =k^2 + \frac{1}{\pi} \int_{-B}^{B} \frac{2c\epsilon_{n}
(\lambda)d \lambda }{c^2+4(k-\lambda)^2},  \label{pp1} \\
&&\epsilon_n(\lambda)= \frac{1}{\pi} \int_{-Q}^{Q}
\frac{2nc\epsilon (k)d k }{c^2+4(\lambda-k)^2}- \frac{1}{\pi}
\int_{-B}^{B} A_{1n}(\lambda-\lambda^{\prime})
\epsilon_1(\lambda^{\prime})d \lambda^{\prime} \nonumber \\
&&\quad\quad\quad  +\frac{1}{\pi} \int_{-D}^{D}
\frac{2nc\epsilon_1^{(1)}( \lambda^{(1)} )d \lambda^{(1)}
}{n^2c^2+4(\lambda-\lambda^{(1)})^2},
\label{pp2} \\
&&\epsilon_1^{(1)}(\lambda^{(1)}) =\frac{1}{\pi} \int_{-B}^{B} \frac{2c
\epsilon_{1} (\lambda)d \lambda }{c^2+4(\lambda^{(1)}-\lambda)^2}.
\label{pp3}
\end{eqnarray}
The dress energies at zero temperature are calculated from the thermodynamic
Limit. The detailed derivations and the further explanations can be found in
the next section.

\subsection{Spin wave excitation}

The second class excitation is flipping one \textquotedblleft spin", which
means add two holes in the distribution of the quantum integer series $%
\{J_{\alpha }^{(1)}\}$. Then the quantum numbers change from integer to
half-odd-integer or vice versa. We denote the spectral parameters
corresponding to the missing integers in the $\{J_{\alpha }^{(1)}\}$ by $%
\lambda _{1}^{h}$ and $\lambda _{2}^{h}$. In the thermodynamic
limit we obtain following coupled integral equations for a state
with two holes
\begin{figure}[ht]
\begin{center}
\includegraphics[height=5.2cm,width=6.3cm]{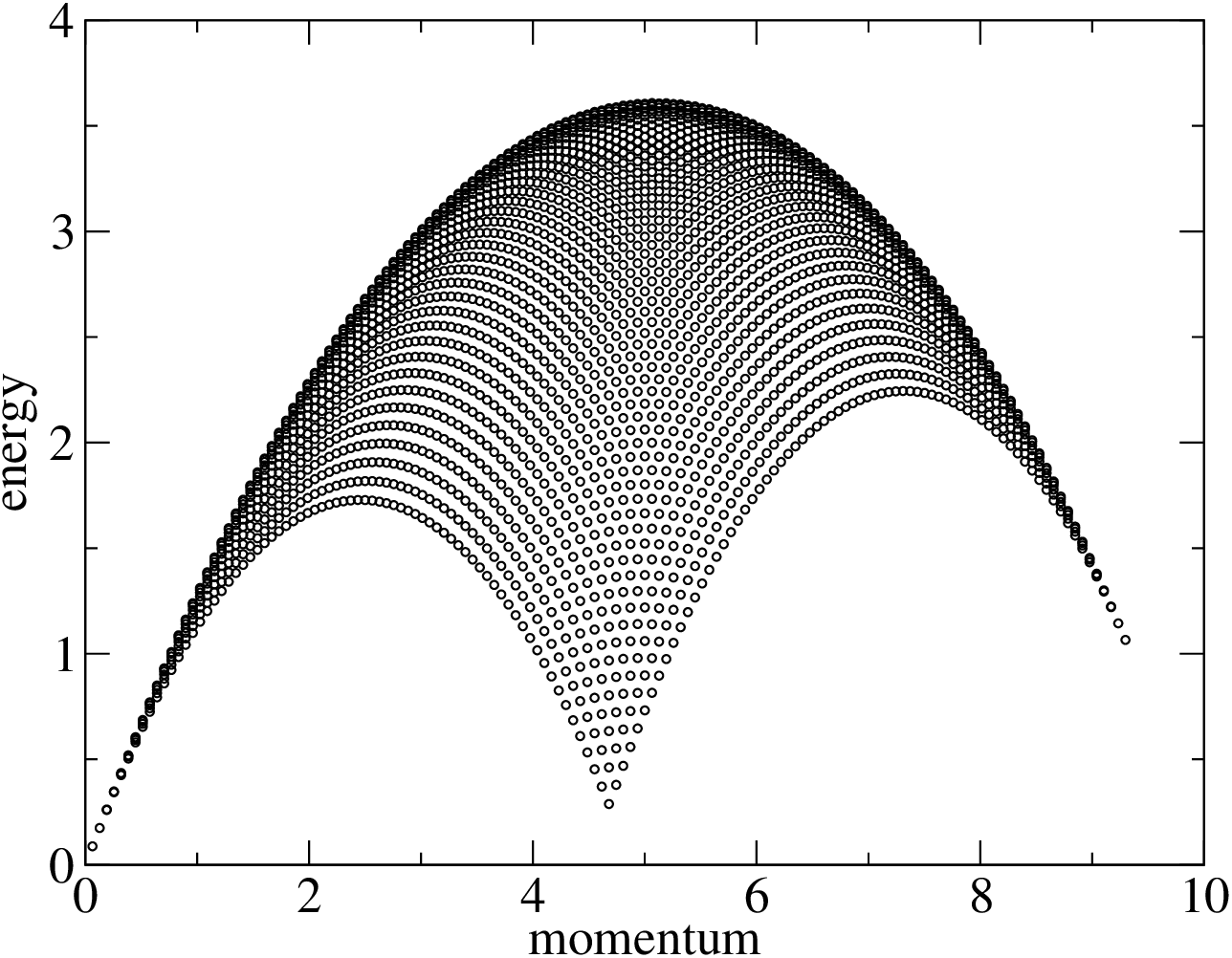}\quad\quad
\includegraphics[height=5.2cm,width=6.3cm]{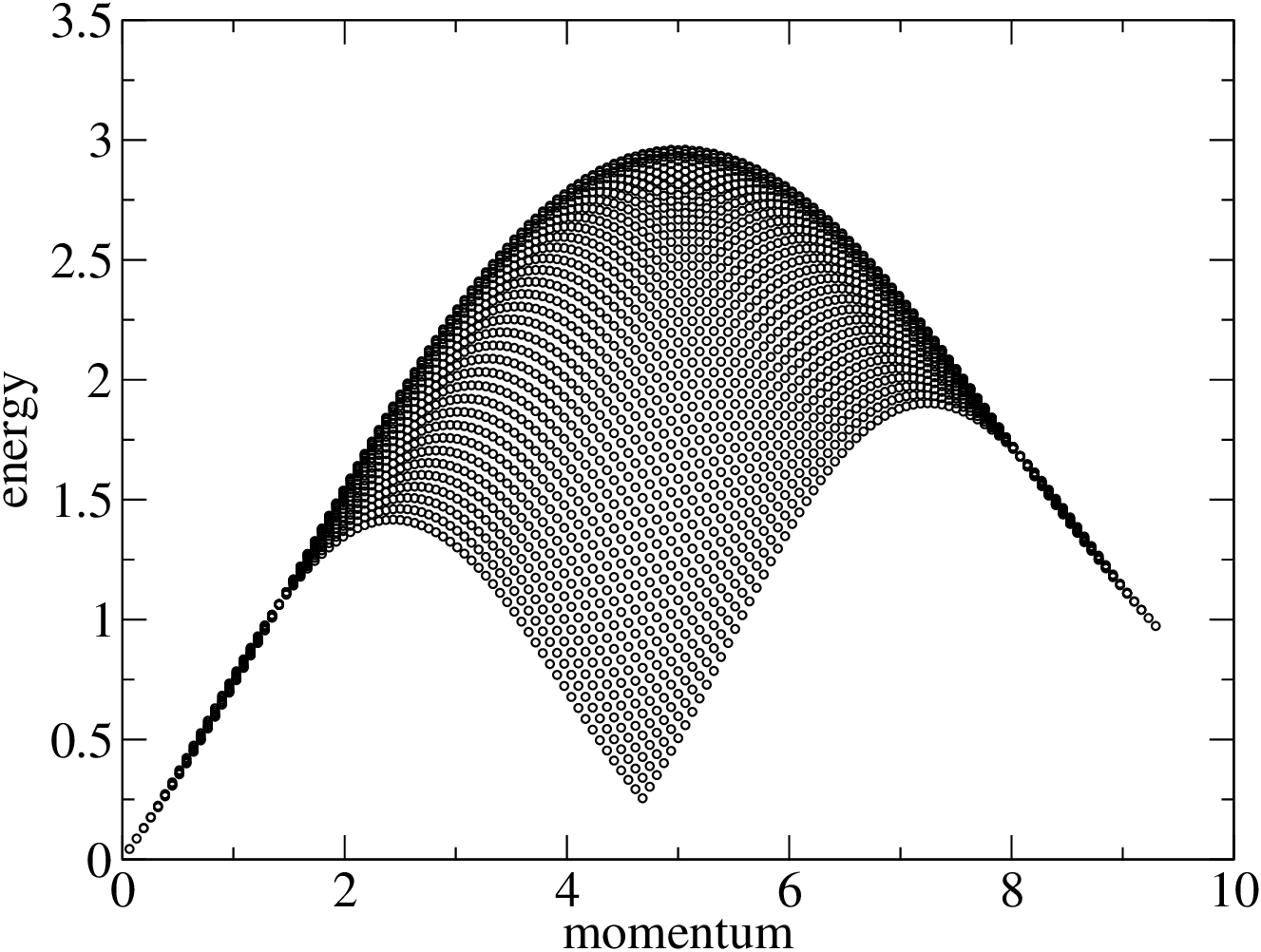}
\end{center}
\caption{The spinon-spinon excitation spectra calculated for
$L=98, N_{f_1}=N_{f_2}=25, N_{b_1}=48$, and the coupling $c=1.0$
(LEFT) and $c=10$ (RIGHT).} \label{fig3}
\end{figure}
\begin{eqnarray}
&&\rho (k)=\frac{1}{2\pi }+\frac{1}{\pi }\int_{-B}^{B}\frac{2c\rho
_{1}(\lambda )d\lambda }{c^{2}+4(k-\lambda )^{2}},
\nonumber \\
&&\rho _{1}(\lambda )=\frac{1}{\pi }\int_{-Q}^{Q}\frac{2c\rho (k)dk}{%
c^{2}+4(\lambda -k)^{2}}-\frac{1}{\pi }\int_{-B}^{B}\frac{c\rho
_{1}(\lambda ^{\prime })d\lambda ^{\prime }}{c^{2}+(\lambda
-\lambda ^{\prime })^{2}} \nonumber \\
&&\quad \quad \quad +\frac{1}{\pi }\int_{-D}^{D}\frac{2c\rho ^{(1)}(\lambda
^{(1)})d\lambda ^{(1)}}{c^{2}+4(\lambda -\lambda ^{(1)})^{2}}-\frac{1}{L}%
\sum_{j=1}^{2}\delta (\lambda -\lambda _{j}^{h}),  \nonumber \\
&&\rho ^{(1)}(\lambda ^{(1)})=\frac{1}{\pi }\int_{-B}^{B}\frac{2c\rho
_{1}(\lambda )d\lambda }{c^{2}+4(\lambda ^{(1)}-\lambda )^{2}}.  \label{sba}
\end{eqnarray}%

The differences of the densities of states and the bare energies are defined
as before. The corrections to the densities at present case are
\begin{eqnarray}
\varphi_1=0, \quad \varphi_2=-\sum_{j=1}^2 \delta(\lambda-\lambda_j^h),
\quad \varphi_3=0.
\end{eqnarray}
We obtain the excitation energy as
\begin{eqnarray}
E_{ex}=E-E_{GS}=\sum_{\alpha=1}^{3}\int_{-a_{\alpha}}^{a_{\alpha}}
\varepsilon_{\alpha}(\mu)\varphi_{\alpha}d\mu
=-\epsilon(\lambda_1^h)-\epsilon(\lambda_2^h).  \label{151}
\end{eqnarray}
This excitation is gapless. The total $S=1$ and the total
$S^z=1$, thus it is a spin triplet excitation.

\begin{figure}[ht]
\begin{center}
\includegraphics[height=6cm,width=8cm]{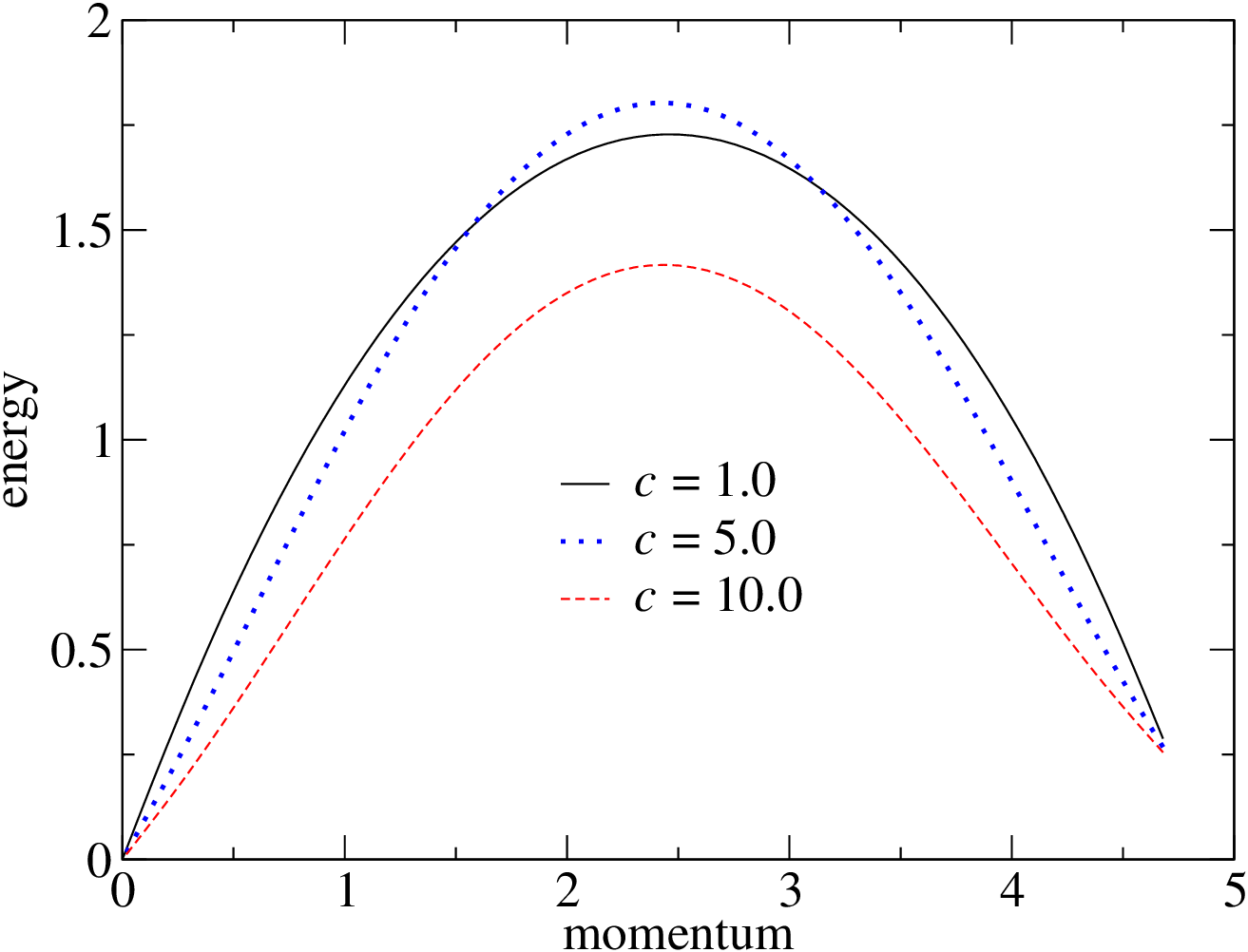}
\end{center}
\caption{The dispersion relation of the spinon excitation.}
\label{fig4}
\end{figure}
Another spin excitation is the spin singlet excitation. This case
means digging two holes in the quantum number sequence $\{
J_{\alpha}^{(1)}\}$ and
constructing one $\lambda$-string of length $2$ (with the quantum number $%
J_{\alpha}^{(2)}$). Thus in the rapidities $\{\lambda\}$, two of them form a
2-string ($\lambda_1^{(2)} \pm ic/2$) and the rest are real. The string
center $\lambda_1^{(2)}$ is determined by
\begin{eqnarray}
\sum_{j=1}^{N} \Xi_{1}\left(\lambda_1^{(2)}-k_j\right) =2\pi
J_{\alpha}^{(2)} +\sum_{\beta=1}^{N_{b_1}+N_{f_2}-2}\Xi_{1/2}
\left( \lambda_1^{(2)}-\lambda_{\beta}^1\right)
+\sum_{\gamma=1}^{N_{b_1}}\Xi_{1}
\left(\lambda_{\gamma}^{(1)r}-\lambda_{1}^{(2)}\right),
\end{eqnarray}
If the system-size keeps finite, the Bethe ansatz equations can be
solved numerically. The excitation spectra and dispersion
relation are shown in Figs. \ref{fig3} and \ref{fig4},
respectively. If the system-size tends to infinity, we obtain the
following coupled integral equations for a state with two holes
and one 2-string
\begin{eqnarray}
&&\rho(k) =\frac{1}{2\pi} +\frac{1}{\pi} \int_{-B}^{B} \frac{2c\rho_{1}
(\lambda)d \lambda }{c^2+4(k-\lambda)^2} +\frac{1}{\pi L}\frac{c}{%
c^2+(k-\lambda)^2},  \nonumber \\
&&\rho_1(\lambda)= \frac{1}{\pi} \int_{-Q}^{Q} \frac{2c\rho (k)d k }{%
c^2+4(\lambda-k)^2}- \frac{1}{\pi} \int_{-B}^{B} \frac{c\rho_{1}
(\lambda^{\prime})d \lambda^{\prime} }{c^2+(\lambda-\lambda^{\prime})^2} +%
\frac{1}{\pi} \int_{-D}^{D} \frac{2c\rho^{(1)}( \lambda^{(1)} )d
\lambda^{(1)} }{c^2+4(\lambda-\lambda^{(1)})^2}  \nonumber \\
&& \quad\quad\quad -\frac{1}{ L}\sum_{j=1}^2\delta(\lambda-\lambda_j^h) -
\frac{1}{\pi L}\left(\frac{2c}{c^2+4(\lambda-\lambda_1^{(2)})^2}+ \frac{6c}{%
9c^2+4(\lambda-\lambda_1^{(2)})^2}\right),  \nonumber \\
&&\rho^{(1)}(\lambda^{(1)}) =\frac{1}{\pi} \int_{-B}^{B} \frac{2c\rho_{1}
(\lambda)d \lambda }{c^2+4(\lambda^{(1)}-\lambda)^2}+\frac{1}{\pi L}\frac{c}{%
c^2+(\lambda_1^{(2)}-\lambda)^2}.  \label{2sba}
\end{eqnarray}
The corrections to the densities of states at present case are
\begin{eqnarray}
&& \varphi_1= \frac{1}{\pi}\frac{c}{c^2+(k-\lambda)^2}, \quad
\varphi_3=\frac{1}{\pi }\frac{c}{c^2+(\lambda_1^{(2)}-\lambda)^2}, \nonumber \\
&&\varphi_2= -\sum_{j=1}^2\delta(\lambda-\lambda_j^h) - \frac{1}{\pi }\left(%
\frac{2c}{c^2+4(\lambda-\lambda_1^{(2)})^2}+ \frac{6c}{9c^2+4(\lambda-%
\lambda_1^{(2)})^2}\right).
\end{eqnarray}
The excitation energy is $
E_{ex}=\sum_{\alpha=1}^{3}\int_{-a_{\alpha}}^{a_{\alpha}}
\varepsilon_{\alpha}(\mu)\varphi_{\alpha}d\mu$.
\begin{figure}[ht]
\begin{center}
\includegraphics[height=5.2cm,width=6.3cm]{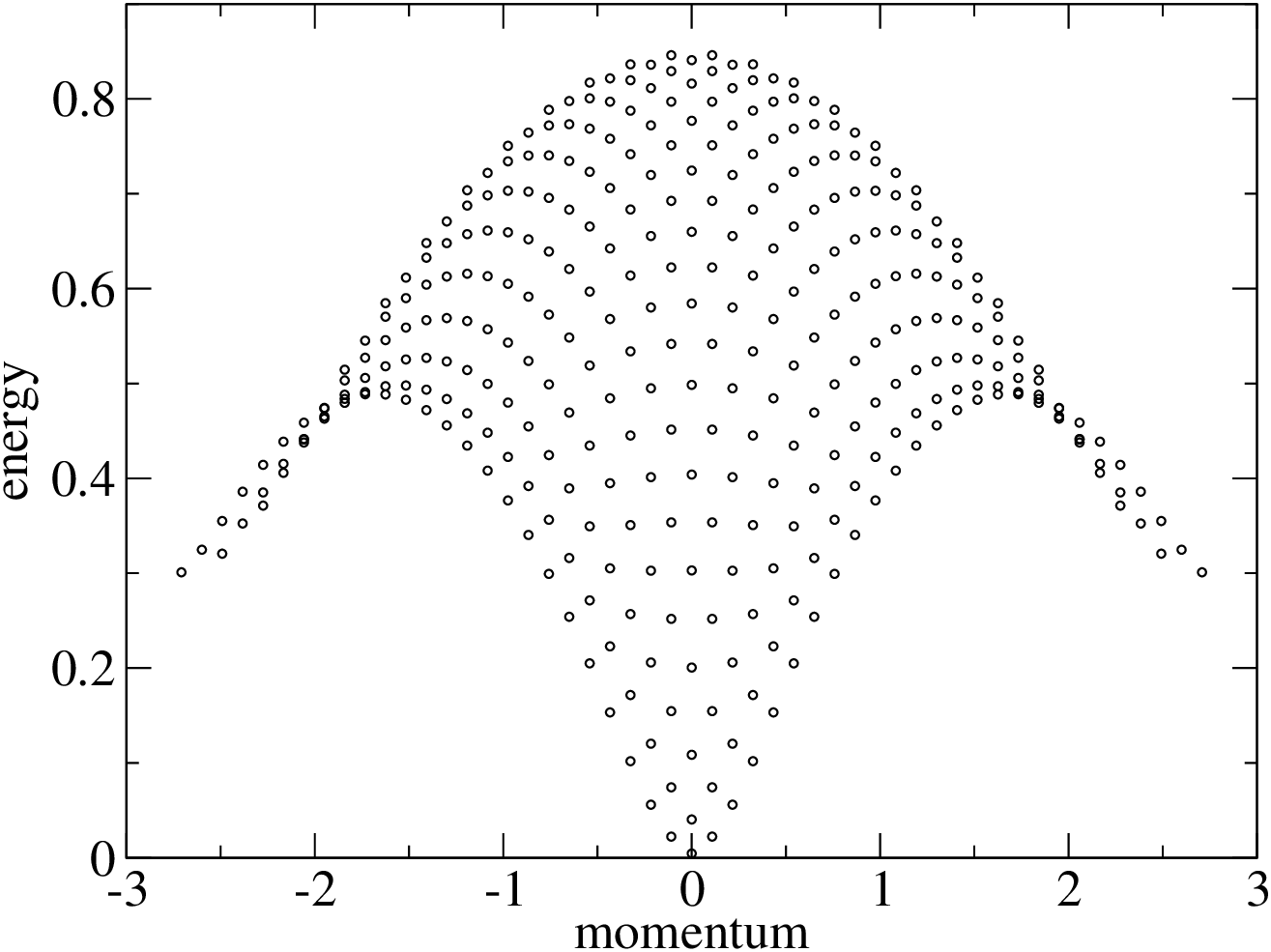}\quad\quad
\includegraphics[height=5.2cm,width=6.3cm]{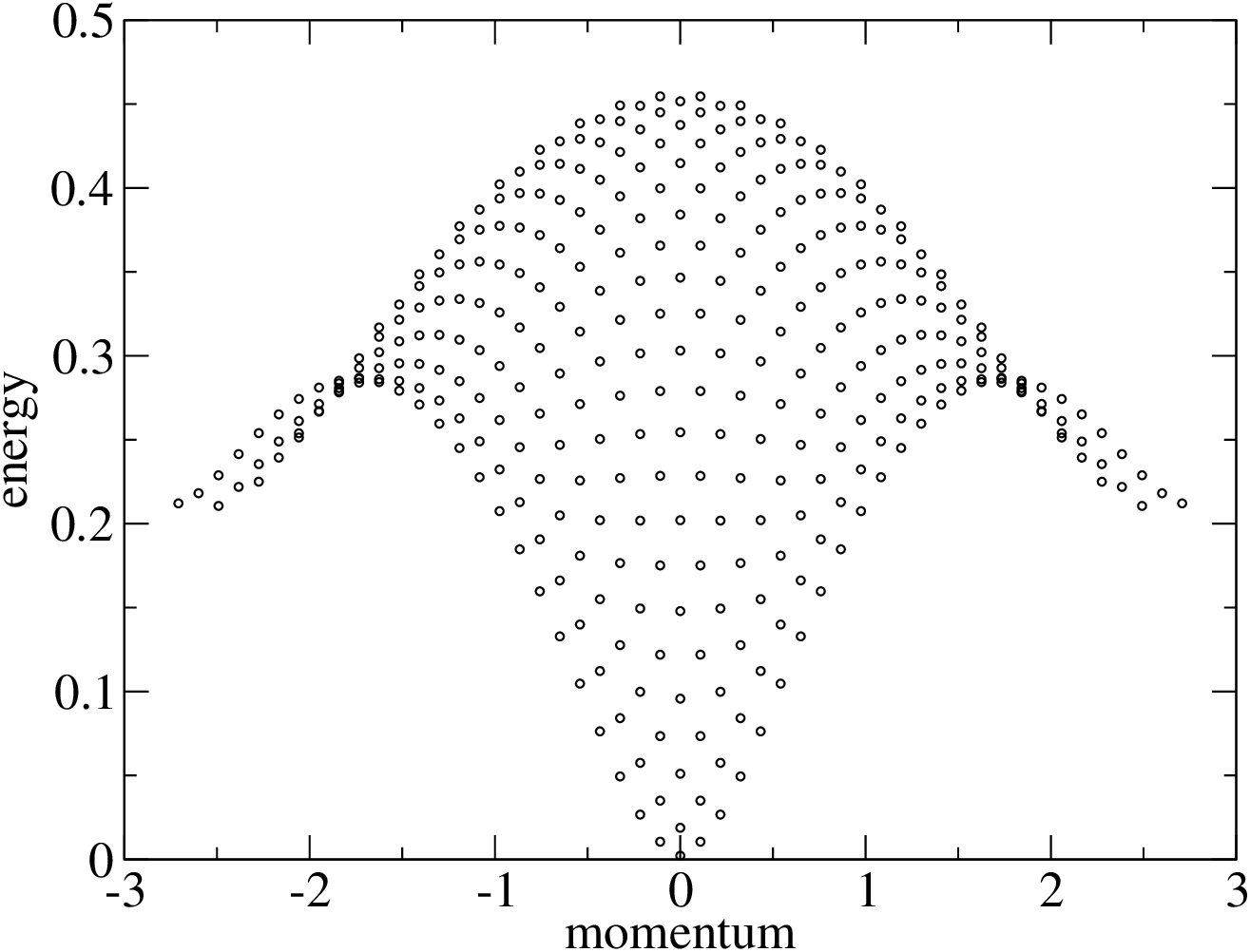}
\end{center}
\caption{The isospinon-isospinon excitation spectra calculated for $L=98, N_{%
\mathrm{f}}=50, N_{\mathrm{b}}=48$, and the coupling $c=1.0$ (LEFT) and $%
c=10 $ (RIGHT).} \label{fig5}
\end{figure}
In order to calculate the excitation energy, we recall the definition (\ref%
{pp2}) of the dressed energy $\epsilon_n(\lambda)$ of $\lambda$-string with
the length $n$. It is easy to proved that $\epsilon_2(\lambda)=0$ for the
present case, then we obtain the excitation energy as
\begin{eqnarray}
E_{ex}=-\epsilon(\lambda_1^h)-\epsilon(\lambda_2^h).  \label{pp}
\end{eqnarray}
We see that the excitation energy (\ref{pp}) is the same as that of the
excitation with only two holes and no strings.

\subsection{Isospin excitation}

The third excitation is replacing $p$ holes in the ground state
distribution of the quantum number series $\{J_{\alpha}^{(2)}\}$ by a $\lambda^{(2)}$%
-string with the length $p$, $\lambda_{j}^{(2)}= \kappa + ic(p+1-2j)/2$,
where $\kappa$ is a real number and $j=1,\cdots, p$. The $p$-string solution
describes an excitation of $p$ particles bound state. This excitation is
spinless. The quantum numbers $I$ and $J_{\alpha}$ are the same as that at
the ground state, while the quantum number $J_{\alpha}^{(1)}$ jumps from
half-odd integer to integer. The allowed range of integer for the $%
\lambda^{(2)}$-string is $|J^{(2)}|\leq \frac{1}{2}(N_{b_1}-2p)$. The center
$\kappa$ of the $\lambda^{(2)}$-string is determined by
\begin{eqnarray}
2\pi J^{(2)p}=\sum_{\gamma=1}^{N_{b_1}} \Xi_{p/2}
\left(\kappa-\lambda_{\gamma}^{(1)}\right).
\end{eqnarray}
Please see the next section for further explanations. The
densities of states at this excitation are
\begin{eqnarray}
&&\rho(k) =\frac{1}{2\pi} +\frac{1}{\pi} \int_{-B}^{B} \frac{2c\rho_{1}
(\lambda)d \lambda }{c^2+4(k-\lambda)^2},  \nonumber \\
&&\rho_1(\lambda)= \frac{1}{\pi} \int_{-Q}^{Q} \frac{2c\rho (k)d k }{%
c^2+4(\lambda-k)^2}- \frac{1}{\pi} \int_{-B}^{B} \frac{c\rho_{1}
(\lambda^{\prime})d \lambda^{\prime}
}{c^2+(\lambda-\lambda^{\prime})^2} + \frac{1}{\pi} \int_{-D}^{D}
\frac{2c\rho^{(1)}( \lambda^{(1)} )d
\lambda^{(1)} }{c^2+4(\lambda-\lambda^{(1)})^2},  \nonumber \\
&&\rho^{(1)}(\lambda^{(1)}) =\frac{1}{\pi} \int_{-B}^{B} \frac{2c\rho_{1}
(\lambda)d \lambda }{c^2+4(\lambda^{(1)}-\lambda)^2} - \frac{1}{\pi L}\frac{%
2pc}{p^2c^2+4(\lambda^{(1)}-\kappa)^2}.
\end{eqnarray}
\begin{figure}[ht]
\begin{center}
\includegraphics[height=6cm,width=8cm]{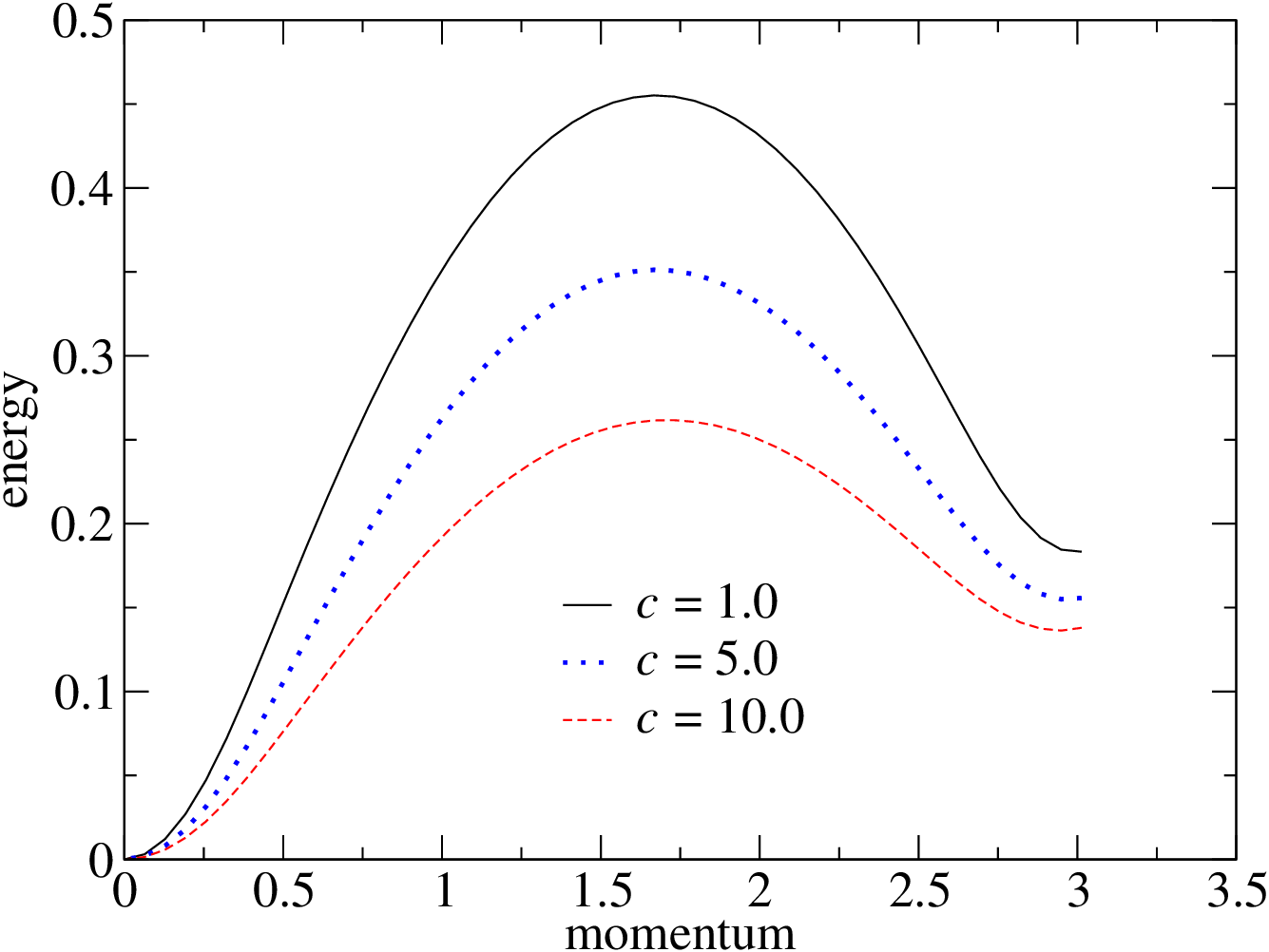}
\end{center}
\caption{The dispersion relation of the elementary excitation:
isospinon.} \label{fig6}
\end{figure}
Thus the corrections to the densities are
\begin{eqnarray}
\varphi_1= \varphi_2=0, \quad \varphi_3=- \frac{1}{\pi}\frac{2pc}{%
p^2c^2+4(\lambda^{(1)}-\kappa)^2}.
\end{eqnarray}
Using the similar method, we obtain the excitation energy as
\begin{eqnarray}
E_{ex}=- \frac{1}{\pi}\int_{-B}^{B} \frac{2c}{c^2+4(\lambda^{(1)}-\kappa)^2}%
\epsilon^{(1)}(\lambda^{(1)})d\lambda^{(1)}= \epsilon_p^{(2)}(\kappa).
\end{eqnarray}

Another isospin excitation is the $\lambda^{(1)} - \lambda^{(2)}$-string
excitation. The quantum number series $\{I_j\}$, $\{J_{\alpha}\}$ and $%
\{J_1^{(1)}\}$ are filled symmetrically around the zero in this excitation.
In the rapidities $\{ \lambda^{(1)} \}$, two of them form the $%
\lambda^{(1)}-\lambda^{(2)}$-string, $\lambda^{(1)}=\lambda^{(2)}\pm ic/2 $
and the rest are real. The allowed range of integer for the $%
\lambda^{(1)}-\lambda^{(2)}$-string is $|J_2^{(1)}|\leq \frac{1}{2}%
(N_{b_1}-1)$. The center of the $\lambda^{(1)}-\lambda^{(2)}$-string $%
\kappa=\lambda^{(2)}$ is determined by
\begin{eqnarray}
2\pi J_2^{(1)}=\sum_{\alpha=1}^{N_{b_1}+N_{f_2}} \Xi_{1}
\left(\kappa-\lambda_{\alpha}\right) - \sum_{\gamma=1}^{N_{b_1}-1} \Xi_{1/2}
\left(\kappa-\lambda_{\gamma}^{(1)}\right).
\end{eqnarray}
The Bethe ansatz equations and energy spectrum for the finite system-size
case can be solved numerically. From that, we obtain the isospinon-isospinon
excitation spectra and corresponding dispersion relation, which are shown in
Figs. \ref{fig5} and \ref{fig6}, respectively. Comparing Figs. \ref{fig4}
and \ref{fig6}, we see that the excitation spectra of spinon is linear while
that of isospinon is quadratic for the small momentum.

If the system-size tends to infinity, we obtain the densities of states at
this excitation as
\begin{eqnarray}
&&\rho(k) =\frac{1}{2\pi} +\frac{1}{\pi} \int_{-B}^{B} \frac{2c\rho_{1}
(\lambda)d \lambda }{c^2+4(k-\lambda)^2},  \nonumber \\
&&\rho_1(\lambda)= \frac{1}{\pi} \int_{-Q}^{Q} \frac{2c\rho (k)d k }{%
c^2+4(\lambda-k)^2}- \frac{1}{\pi} \int_{-B}^{B} \frac{c\rho_{1}
(\lambda^{\prime})d \lambda^{\prime}
}{c^2+(\lambda-\lambda^{\prime})^2} \nonumber
\end{eqnarray}
\begin{eqnarray}
&&\quad\quad\quad + \frac{1}{\pi} \int_{-D}^{D}
\frac{2c\rho^{(1)}( \lambda^{(1)} )d
\lambda^{(1)} }{c^2+4(\lambda-\lambda^{(1)})^2} +\frac{1}{\pi L} \frac{c}{%
c^2+(\lambda-\kappa)^2},  \nonumber \\
&&\rho^{(1)}(\lambda^{(1)}) =\frac{1}{\pi} \int_{-B}^{B} \frac{2c\rho_{1}
(\lambda)d \lambda }{c^2+4(\lambda^{(1)}-\lambda)^2} - \frac{1}{\pi L}\frac{%
2c}{c^2+4(\lambda^{(1)}-\kappa)^2}.
\end{eqnarray}
Thus the corrections to the densities are
\begin{eqnarray}
\varphi_1= 0, \quad \varphi_2= \frac{1}{\pi} \frac{c}{c^2+(\lambda-\kappa)^2}%
, \quad \varphi_3=- \frac{1}{\pi}\frac{2c}{c^2+4(\lambda-\kappa)^2}.
\end{eqnarray}
Using the similar method, we obtain the excitation energy as
\begin{eqnarray}
E_{ex}(\kappa) =\frac{1}{\pi}\int_{-D}^{D} \frac{c}{c^2+(\lambda-\kappa)^2}
\epsilon_{1}(\lambda)d\lambda -\frac{1}{\pi}\int_{-B}^{B} \frac{2c}{%
c^2+4(\lambda^{(1)}-\kappa)^2}\epsilon^{(1)}(\lambda^{(1)})d\lambda^{(1)}=
\epsilon_2^{(1)}(\kappa).
\end{eqnarray}

\section{Thermodynamics of the system}

The finite-temperature properties of the system can be studied based on the
solutions of the Bethe ansatz equations in the thermodynamic limit. The
solutions of the Bethe ansatz equations are a little bit complicated.
Besides real solutions, the Bethe ansatz equations also have complex
solutions. Generally, the complex solutions are determined by the poles or
zeros of the Bethe ansatz equations in the thermodynamic limit. Which gives
us a hint to determine the structures of the solutions of the Bethe ansatz
equations. In the following, we only consider the case $c>0$, because the
bosons with attractive interactions ($c<0$) do not have the thermodynamics
\cite{takahashi}. After some analysis, we find that the structures of the
solutions of the Bethe ansatz equations (\ref{2bea1}) - (\ref{2bea4}) are:
(1) All the momentums $k_j$ are real. (2) The rapidities $\lambda$ form the $%
m$-string, $\lambda_{\alpha}^{mj}=\lambda_{\alpha}^m + (m+1-2j) ic /2
+o(e^{-\delta L})$, where $\lambda_{\alpha}^m $ are real and $j=1,\cdots, m$%
. (3) Some rapidities $\lambda^{(1)}=\lambda^{(1)r}$ are real. (4) Some
rapidities $\lambda^{(1)}$ form the $\lambda^{(1)}-\lambda^{(2)}$-strings, $%
\lambda_{\gamma}^{(1)} =\lambda^{(1)s} \pm ic/2 +o(e^{-\delta L})$ and $%
\lambda_{\gamma}^{(2)}=\lambda^{(1)s}$, where $\lambda^{(1)s}$ are real. (5)
Some rapidities $\lambda^{(2)}$ are real, which are the real part of the $%
\lambda^{(1)}-\lambda^{(2)}$-strings. (6) Some $\lambda^{(2)}$ form the $p$%
-strings, $\lambda_{\nu}^{(2)pl}=\lambda_{\nu}^{(2)p} +
(p+1-2l)ic/2 +o(e^{-\delta L})$, where $\lambda_{\nu}^{(2)p}$ are
real and $l=1,\cdots, p$.

At the temperature $T$, the system (1) arrive at the thermal
equilibrium. We denote the density of momentum $k$ by $\rho(k)$, the density of $\lambda$%
-strings with length $n$ by $\rho_n(\lambda)$, the density of real rapidity $%
\lambda^{(1)}$ by $\rho_1^{(1)}(\lambda^{(1)})$, the density of $%
\lambda^{(1)}-\lambda^{(2)}$-strings by $\rho_2^{(1)}(\lambda^{(1)})$ and
the density of $\lambda^{(2)}$-strings by $\rho_p^{(2)}(\lambda^{(2)})$.
Meanwhile, the notations $\rho^h(k)$, $\rho_n^h(\lambda)$, $%
\rho_1^{(1)h}(\lambda^{(1)})$, $\rho_2^{(1)h}(\lambda^{(1)})$ and $%
\rho_p^{(2)h}(\lambda^{(2)})$ represents the densities of
corresponding holes. From the Bethe ansatz equations
(\ref{2bea1}) - (\ref{2bea4}), these densities should satisfy
\begin{eqnarray}
&& \rho(k)+\rho^h(k)=\frac{1}{2\pi}+\sum_{n=1}^{\infty}a_n*\rho_n(k),
\nonumber \\
&&\rho_n(\lambda)+\rho_n^h(\lambda)=a_n*\rho(\lambda)
-\sum_{m=1}^{\infty}A_{n,m}*\rho_{m}(\lambda)
+(A_{n,1}-\delta_{n,1})*\rho_2^{(1)}(\lambda) +a_n*\rho_1^{(1)}(\lambda),
\nonumber \\
&& \rho_1^{(1)}(\lambda^{(1)})+\rho_1^{(1)h}(\lambda^{(1)})=
\sum_{n=1}^{\infty}a_n*\rho_{n}(\lambda^{(1)})
-\sum_{p=1}^{\infty}a_p*\rho_{p}^{(2)}(\lambda^{(1)})
-a_1*\rho_{2}^{(1)}(\lambda^{(1)}),  \nonumber \\
&&\rho_2^{(1)}(\lambda^{(1)})+\rho_2^{(1)h}(\lambda^{(1)})
=\sum_{n=1}^{\infty}(A_{n,1}-\delta_{n,1})*\rho_{n}(\lambda^{(1)})
-a_1*\rho_{1}^{(1)}(\lambda^{(1)}) -a_2*\rho_{2}^{(1)}(\lambda^{(1)}),
\nonumber \\
&&\rho_p^{(2)}(\lambda^{(2)})+\rho_p^{(2)h}(\lambda^{(2)})=
-\sum_{q=1}^{\infty}A_{p,q}*\rho_{q}^{(2)}(\lambda^{(2)})
+a_p*\rho_{1}^{(1)}(\lambda^{(2)}),
\end{eqnarray}
where the integral operators $a_n(x)$ and $A_{nm}(x)$ are
\begin{eqnarray}
&&a_n(x)=\frac{1}{\pi}\frac{2nc}{n^2c^2+4x^2}, \quad a_0(x)=\delta(x),
\nonumber \\
&&A_{n,m}(x)=\left\{%
\begin{array}{cc}
a_{|n-m|}(x)+2a_{|n-m|+2}(x)+\cdots+2a_{n+m-2}(x)+a_{n+m}(x) &
\mathrm{if
} \quad n\neq m, \\
2a_{2}(x)+2a_{4}(x)+\cdots+2a_{2n-2}(x)+a_{2n}(x), & \mathrm{if}
\quad n= m.
\end{array}%
\right.
\end{eqnarray}
The convolution is defined as $a*b(x)=\int a(x-y)b(y)dy$.

The Gibbs free energy at a given temperature $T$, chemical potential $A$ and
external magnetic field $h$ is given by
\begin{eqnarray}
G(T,A,h)=E -A N - h (M^f+M^b)-T \mathcal{S}.
\end{eqnarray}
The densities of energy and particles number are $E/L=
\int_{-Q}^{Q} k^2 \rho(k) dk$, $ N/L=\int_{-Q}^{Q} \rho(k) dk$.
The magnetization for fermions and that for bosons are
\begin{eqnarray}
&&M^f=N_{f_1}-N_{f_2}=-M_1^{(1)}-2 M_2^{(1)} +2\sum_{n=1}^{\infty} nM_n -N,
\\
&&M^b=g(N_{b_1}-N_{b_2})=2g\sum_{p=1}^{\infty}pM_p^{(2)}-gM_1^{(1)},
\end{eqnarray}
where $g$ is the Landau $g$-factor. The entropy of the system is
$\mathcal{S}/{L}=\int \left[(\rho+\rho^h)\ln(\rho+\rho^h)\right.$
$\left. -\rho\ln\rho-\rho^h\ln\rho^h \right]dk
+\sum_{n=1}^{\infty} \int\left[(\rho_n+\rho_n^h)
\ln(\rho_n+\rho_n^h) -\rho_n\ln\rho_n -\rho_n^h\ln\rho_n^h
\right]d\lambda +  \int\left[(\rho_1^{(1)} +\rho_1^{(1)h})
\right.$\\ $\left. \ln(\rho_1^{(1)} +\rho_1^{(1)h}) -\rho_1^{(1)}
\ln\rho_1^{(1)} -\rho_1^{(1)h}\ln\rho_1^{(1)h}\right]
d\lambda^{(1)} + \int \left[(\rho_2^{(1)}
+\rho_2^{(1)h})\ln(\rho_2^{(1)} +\rho_2^{(1)h}) -\rho_2^{(1)}
\ln\rho_2^{(1)}- \right.$\\
$\left.\rho_2^{(1)h}\ln\rho_2^{(1)h}\right]d\lambda^{(1)} +
\sum_{p=1}^{\infty} \int\left[(\rho_p^{(2)} +\rho_p^{(2)h})
\ln(\rho_p^{(2)}+\rho_p^{(2)h}) -\rho_p^{(2)}\ln\rho_p^{(2)}
-\rho_p^{(2)h}\ln\rho_p^{(2)h} \right]d\lambda^{(2)}$.

For convenience, we introduce following notations $\xi=
\rho^h/\rho,$ $\alpha_n=\rho_n^h/\rho_n$, $
\beta_1=\rho_1^{(1)h}/\rho_1^{(1)}$, $\beta_2=\rho_2^{(1)h}
/\rho_2^{(1)}$ and $\gamma_p=\rho_p^{(2)h}/\rho_p^{(2)}$. At the
thermodynamic equilibrium state, the free energy must be
minimized. From the variation of free energy is zero, we obtain
following coupled non-linear integration equations
\begin{eqnarray}
&&\ln \xi(k) =\frac{k^2-A+h}{T} -\sum_{n=1}^{\infty} a_n * \ln
(1+\alpha_n^{-1}(k)),  \label{tbae1} \\
&&\ln \alpha_n(\lambda)= - \frac{2nh}{T} +\sum_{m=1}^{\infty} A_{m,n} * \ln
(1+\alpha_m^{-1}(\lambda)) -a_{n} * \ln (1+\xi^{-1}(\lambda))  \nonumber \\
&& \hspace{2cm} - a_{n} * \ln (1+\beta_1^{-1}(\lambda))-
(A_{n,1}-\delta_{n,1}) * \ln (1+\beta_2^{-1}(\lambda)),  \label{tbae2} \\
&&\ln \beta_1(\lambda^{(1)})= \frac{(1+g)h}{T} - \sum_{n=1}^{\infty} a_n *
\ln (1+\alpha_n^{-1}(\lambda^{(1)}))  \nonumber \\
&& \hspace{2cm} + a_{1} * \ln (1+\beta_2^{-1}(\lambda^{(1)}))-
\sum_{p=1}^{\infty} a_{p} * \ln (1+\gamma_p^{-1}(\lambda^{(1)})),
\label{tbae3} \\
&&\ln \beta_2(\lambda^{(1)})= \frac{2h}{T} - \sum_{n=1}^{\infty}
(A_{n,1}-\delta_{n,1}) * \ln (1+\alpha_n^{-1}(\lambda^{(1)}))  \nonumber \\
&& \hspace{2cm}+ a_{2} * \ln (1+\beta_2^{-1}(\lambda^{(1)}))+ a_1 * \ln
(1+\beta_1^{-1}(\lambda^{(1)})),  \label{tbae4} \\
&&\ln \gamma_p(\lambda^{(2)})= - \frac{2pgh}{T} +
\sum_{q=1}^{\infty} A_{q,p} * \ln
(1+\gamma_p^{-1}(\lambda^{(2)})) + a_{p} * \ln
(1+\beta_1^{-1}(\lambda^{(2)})),  \label{tbae5}
\end{eqnarray}
where $*$ denotes the convolution.

\section{Special limits}

In principle, the thermodynamic Bethe ansatz equations can not be solved
analytically. One has to use the numerical simulations or the approximate
methods. However, at some special limit cases, the thermodynamic Bethe
ansatz equations can be solved exactly.

\subsection{Zero temperature limit}

We first consider the zero temperature limit. The dressed
energies $\epsilon, \epsilon_n, \epsilon_1^{(1)},
\epsilon_2^{(1)}$ and $\epsilon_p^{(2)}$ are defined as
$\xi=\exp(\epsilon/T)$, $\alpha_n=\exp\left(
\epsilon_n/T\right)$, $\beta_1=\exp\left(\epsilon_1^{(1)}/T
\right)$, $ \beta_2=\exp\left(\epsilon_2^{(1)}/T\right)$ and $
\gamma_p=\exp\left(\epsilon_p^{(2)}/T\right)$. Substituting them
into the thermodynamic Bethe ansatz Eqs. (\ref{tbae1}) -
(\ref{tbae5}) and letting the temperature tends to zero, we
obtain the analytic formulas for the dressed energies
\begin{eqnarray}
&&\epsilon(k) =k^2 + \frac{1}{\pi} \int_{-B}^{B} \frac{2c\epsilon_{n}
(\lambda)d \lambda }{c^2+4(k-\lambda)^2},  \label{dress1} \\
&&\epsilon_n(\lambda)= \frac{1}{\pi} \int_{-Q}^{Q} \frac{2nc\epsilon (k)d k
}{c^2+4(\lambda-k)^2}- \frac{1}{\pi} \int_{-B}^{B}
A_{1,n}(\lambda-\lambda^{\prime}) \epsilon_1(\lambda^{\prime})d
\lambda^{\prime} \nonumber \\
&&\qquad \qquad +\frac{1}{\pi} \int_{-D}^{D}
\frac{2nc\epsilon_1^{(1)}( \lambda^{(1)} )d \lambda^{(1)}
}{n^2c^2+4(\lambda-\lambda^{(1)})^2},
\label{dress2} \\
&&\epsilon_1^{(1)}(\lambda^{(1)}) =\frac{1}{\pi} \int_{-B}^{B} \frac{2c
\epsilon_{1} (\lambda)d \lambda }{c^2+4(\lambda^{(1)}-\lambda)^2},
\label{dress3} \\
&&\epsilon_2^{(1)}(\lambda^{(1)}) =\frac{1}{\pi} \int_{-B}^{B} \frac{c
\epsilon_{1} (\lambda)d \lambda }{c^2+(\lambda^{(1)}-\lambda)^2}-\frac{1}{\pi%
} \int_{-D}^{D} \frac{2c \epsilon_{1}^{(1)} (\lambda)d \lambda }{%
c^2+4(\lambda^{(1)}-\lambda)^2},  \label{dress4} \\
&&\epsilon_p^{(2)}(\lambda^{(2)}) =- \frac{1}{\pi} \int_{-B}^{B} \frac{2pc
\epsilon_{1}^{(1)} (\lambda^{(1)})d \lambda^{(1)} }{c^2+4(\lambda^{(2)}-%
\lambda^{(1)})^2}.
\end{eqnarray}
These equations have been used in calculating the low-lying excitation
energies.

\subsection{High temperature limit}

If the temperature $T$ tends to infinity, the rapidities
$\alpha_n, \beta_1, \beta_2 $ and $\gamma_p$ become constants and
the thermodynamic Bethe ansatz equations become a set of coupled
algebraic equations. The thermodynamic Bethe ansatz Eqs.
(\ref{tbae2}) and (\ref{tbae5}) reads
\begin{eqnarray}
&&\ln \alpha_1 =\frac{1}{2} \ln\left[\frac{1+\alpha_2}{1+\beta_1^{-1}}\right]%
, \quad \ln \alpha_2 =\frac{1}{2}\ln\left[\frac{(1+\alpha_1) (1+\alpha_3)}{%
1+\beta_2^{-1}}\right],  \nonumber \\
&&\ln \alpha_n =\frac{1}{2}\ln\left[(1+\alpha_{n-1})(1+\alpha_{n+1})\right],
\quad n\geq 3,  \nonumber \\
&&\ln \gamma_1 =\frac{1}{2}\ln\left[(1+\gamma_2)(1+\beta_1^{-1})\right],
\quad \ln \gamma_p =\frac{1}{2}\ln\left[(1+\gamma_{p-1})(1+\gamma_{p+1})%
\right], \quad p \geq 2.  \label{therm2}
\end{eqnarray}
The solutions of Eq. (\ref{therm2}) are
\begin{eqnarray}
&& \alpha_n = g^2(n)-1, \quad n\geq 2, \quad \alpha_1=g(2)/f(0),  \nonumber
\\
&& \gamma_p = f^2(p)-1, \quad p\geq 1, \quad \beta_1=1/[f^2(0)-1],
\label{therm3}
\end{eqnarray}
where $g(n)=(a^n-a^{-n})/(a-a^{-1})$, $f(p)=(bd^p-b^{-1}d^{-p})/(
d-d^{-1})$ and the parameters $a, b, d$ are determined by
\begin{eqnarray}
&&\ln \xi= \frac{k^2-A+h}{T} - \sum_{n=1}^{\infty}\ln (1+\alpha_n^{-1}),
\nonumber \\
&&\ln (1+\alpha_1) = -\frac{2h}{T} +
2\sum_{n=1}^{\infty}\ln(1+\alpha_n^{-1})-\ln(1+\xi^{-1})
-\ln(1+\beta_1^{-1})-\ln(1+\beta_2^{-1}),  \nonumber \\
&& \ln \beta_1= \frac{(1+g)h}{T}- \sum_{n=1}^{\infty}\ln (1+\alpha_n^{-1})+
\ln(1+\beta_2^{-1})-\sum_{p=1}^{\infty}\ln (1+\gamma_p^{-1}),  \nonumber \\
&& \ln \beta_2= \frac{2h}{T}- 2\sum_{n=1}^{\infty}\ln (1+\alpha_n^{-1})+ \ln
(1+\alpha_1^{-1})+ \ln(1+\beta_1^{-1})+ \ln(1+\beta_2^{-1}),  \nonumber \\
&&\ln(1+\gamma_1)=-\frac{2gh}{T}+ 2\sum_{p=1}^{\infty}\ln
(1+\gamma_p^{-1})+ \ln(1+\beta_1^{-1}),  \nonumber \\
&&\lim_{n\rightarrow\infty}\frac{\ln\alpha_n}{n}=\frac{2h}{T},
\quad
\lim_{p\rightarrow\infty}\frac{\ln\gamma_p}{p}=\frac{2gh}{T}.
\label{therm5}
\end{eqnarray}

\subsection{Weak coupling limit}

If the coupling constant $c$ tends to zero, the thermodynamic Bethe ansatz
equations can be simplified as
\begin{eqnarray}
&&\ln \xi(k)= \frac{k^2-A}{T} -\frac{1}{2} \ln \left[(1+\alpha_1(k))(1+%
\xi^{-1}(k)) (1+\beta_1^{-1}(k))(1+\beta_2^{-1}(k))\right],  \nonumber \\
&&\ln \alpha_1(\lambda) =\frac{1}{2} \left[\ln (1+\alpha_2(\lambda))-\ln
(1+\beta_1^{-1}(\lambda))\right],  \nonumber \\
&& \ln \alpha_2(\lambda) =\frac{1}{2} \left[\ln(1+\alpha_1(\lambda))+\ln
(1+\alpha_3(\lambda))-\ln(1+\beta_2^{-1}(\lambda))\right],  \nonumber \\
&&\ln \alpha_n(\lambda) =\frac{1}{2}\ln[(1+\alpha_{n-1}(\lambda))(1+%
\alpha_{n+1}(\lambda))], \quad n\geq 3,  \nonumber \\
&&\ln \beta_1(\lambda^{(1)})=-
\frac{1}{2}\ln[(1+\alpha_1(\lambda^{(1)}))
(1+\gamma_1(\lambda^{(1)}))(1+\xi^{-1}
(\lambda^{(1)}))(1+\beta_2^{-1}(\lambda^{(1)}))], \nonumber \\
&&
\ln \beta_2(\lambda^{(1)})=- \ln
[\alpha_1(\lambda^{(1)})(1+\xi^{-1}(\lambda^{(1)}))],  \nonumber \\
&&\ln \gamma_1(\lambda^{(2)}) =\frac{1}{2}\left[\ln(1+\gamma_2(%
\lambda^{(2)}))-\ln (1+\beta_1^{-1}(\lambda^{(2)}))\right],  \nonumber \\
&&\ln \gamma_p(\lambda^{(2)}) =\frac{1}{2}\ln[(1+\gamma_{p-1}(%
\lambda^{(2)})) (1+\gamma_{p+1}(\lambda^{(2)}))], \quad p \geq 2,  \nonumber
\\
&& \lim_{n\rightarrow\infty}\frac{\ln\alpha_n(k)}{n}=\frac{2h}{T}, \quad
\lim_{p\rightarrow\infty}\frac{\ln\gamma_p(\lambda^{(2)})}{p}=\frac{2gh}{T}.
\end{eqnarray}

\subsection{Strong coupling limit}

If the coupling constant $c$ tends to infinity, the momentum rapidities are
completely decoupled with other rapidities. Meanwhile, the solutions for $%
\alpha_n, \beta_1, \beta_2 $ and $\gamma_p$ turn into constants
and are independent of the variables $\lambda, \lambda^{(1)}$ and
$\lambda^{(2)}$. The thermodynamic Bethe ansatz equations can be
solved analytically in this case. The solutions of Eqs.
(\ref{tbae2}) and (\ref{tbae5}) still take the form of
(\ref{therm3}), where the boundary conditions (\ref{therm5})
become $\alpha_1=\beta_2^{-1}$ and $\ln \beta_1=-
\frac{1}{2}\ln(1+\gamma_1)$. Then the analytic solutions of
thermodynamic Bethe ansatz equations are
\begin{eqnarray}
&&\xi= e^{\frac{k^2-A}{T}}\left(2\cosh \frac{gh}{T} +2\cosh\frac{h}{T}%
\right)^{-1}, \quad \alpha_n=\left(\frac{\sinh
\frac{nh}{T}}{\sinh\frac{h}{T}}\right)^{2}-1, \quad n\geq 2,\nonumber \\
&&  \alpha_1=\frac{\cosh\frac{h}{T}}{\cosh\frac{gh}{T}}=\beta_2^{-1},
\quad \beta_1= \frac{1}{4 \cosh^2 \frac{gh}{T}-1}, \quad \gamma_p=\left(\frac{%
\sinh \frac{(p+2)gh}{T}}{\sinh\frac{gh}{T}}\right)^{2}-1, \;
p\geq 1.
\end{eqnarray}

\section{Conclusions}

In summary, we study the exact solutions of 1D mixture of spinor bosons and
spinor fermions with $\delta$-function interactions. The wave function for
the bosonic parties is symmetric while for the fermionic parties is
anti-symmetric. The global wave function of the system is supersymmetric.
After obtaining the two-body scattering matrix by using the coordinate Bethe
ansatz method, we prove that the system is integrable. Then we derive the
energy spectrum and the Bethe ansatz equations with different gradings by
using the graded nest quantum inverse scattering or algebraic Bethe ansatz
method. Based on the solutions of the Bethe ansatz equations, we discuss the
ground state properties of the system. We find that if the interactions are
repulsive, the fermions form spin singlet states and the bosons are
polarized, thus the global ground state is partial polarized. If the
interactions are attractive, the bosons condensed at the zero momentum
point. We discuss the charge-hole excitation, spin wave excitation, isospin
wave excitation and corresponding excitation energies very detailed. We also
obtain the thermodynamic Bethe ansatz equations at finite temperature and
find their analytic solutions at some special limit cases.

\section*{Acknowledgments}

This work is supported by the Earmarked Grant for Research from
the Research Grants Council of HKSAR, China (Project Nos. CUHK
402107 and CUHK 401108), NSF of China, and the national program
for basic research of MOST under Grant No. 2006CB921300. J Cao
acknowledges the financial support from the C N Yang Foundation.

\end{document}